\documentclass[preprint2]{aastex}
\usepackage{graphicx}
\usepackage{epstopdf}

\begin{document}

\title{Magnetic fields in circumstellar envelopes of evolved AGB stars}

\author{G. Pascoli}
\affil{Universit\'e de Picardie Jules Verne, Facult\'e des
Sciences, D\'epartement de Physique, 33 rue Saint-Leu, Amiens, Cedex
80039, France;  pascoli@u-picardie.fr}

\begin{abstract}
In this paper, a time-dependent magnetohydrodynamic model is presented which aimed at understanding the superwind production by an evolved  AGB  star and the consecutive formation of a dense circumstellar envelope around it. We know henceforth from various observations that a large scale  magnetic field, probably toroidal in shape, is duly attested  within these envelopes. Where does this large scale coherent field come  from ? 

\noindent The apparent antinomy between the quasi-round dense circumstellar envelopes and their likely descendants, i.e.  the elongated or bipolar Planetary Nebulae is also questioned. How is the spherical symmetry  broken ?  We suggest in the present model   that  the nebula must effectively appear  round  during the superwind  phase from the  point of view of a distant  observer.  By  contrast  anisotropic structures  are already  appearing  at the same time, but these ones  remain   hidden in the innermost  regions.  We predict  thus  the existence of  a   large bipolar cavity    above the AGB star during the slow superwind  phase.    We then  conjecture that  the  PPNe phase begins when the fast wind  emitted by the core   engulfs  this  cavity and increases  the anisotropy of the distribution of gas.  Thus even though paradoxically enough a beautiful evolved PNe can eventually  emerge   from  a quasi-  round dense circumstellar envelope.

\vspace{5pt}
\noindent Manuscript accepted for publication  in  Publ.Astron.Soc.Pac. (october, 2019)
\end{abstract}

\keywords{-- MHD : dynamo  --  stars:  AGB, Post--AGB, mass loss  --  circumstellar envelopes: magnetic fields,
-- planetary nebulae : morphology}

\section{Introduction}

The origin of magnetic fields in dense circumstellar envelopes  surrounding evolved AGB stars is still today a puzzling problem. The large number of observational studies contrasts with the fact that hitherto solely a few theoretical works are devoted to this topic (Pascoli, 1987, 1992, 1997; Chevalier and Luo, 1994; Matt at al, 2000; Blackman et al, 2001; Garc\'{\i}a-Segura, L\'{o}pez and Franco, 2005; Nordhaus, Blackman and Franck, 2007; Nucci and Busso, 2014). The well-posed problem is the following : the  dense circumstellar envelopes produced  at the tip of the AGB stage  possess weak expansion velocities,  but a contrario,  a large-scale magnetic field is presumably present.   In the first case (weak expansion velocities)  the progenitor must have a very large radius,  which necessarily implies  low rotational velocities;  however in the second case   (presence  of a coherent magnetic field over large  length scales)  the rotation for   the dynamo to be  operating must  be high. How can you solve this dilemma ?

{\raggedright 
After leaving the main sequence low- and intermediate-mass stars, with
an initial mass range of about $0.8 - 7\ M_\odot$,  reach the asymptotic giant branch   (AGB).  The total duration of the AGB phase is of the order of $10^5-10^6$ years.   
The AGB stars are  powered by alternating burning
of H and He in thin shells surrounding their inert  C-O core (Busso, Gallino and  Wasserburg, 1999). Very schematically  this  core is dense and very small with a radius of the order of  $10^9\  cm$.  It is surrounded by a very huge  convective envelope with a radius of the order of  $10^{13}\  cm$. 
Mass-loss rates of AGB stars, determined with various observational methods  are
typically in the range of $10^{-8}–-10^{-5}\  M_\odot\  year^{-1}$ (Habing, 1996). This is the now broadly  accepted  model  of  an archetypical  AGB star   (H\"{o}fner and Olofsson, 2018).
}

\noindent However in the present paper we look at  a short  interval of time  of the full  AGB stage. We are  mostly concerned with the massive wind (the so-called superwind)  produced during  a late period  of the AGB stage, namely    at  the  tip  of the AGB phase  where   mass  losses  as high as $10^{-4}\   M_\odot\  year^{-1}$  or even above, are measured  (using for instance  the  OH 1612 MHz  maser line observations of   OH/IR stars).  This period is  very short, of the order of   $10^4$ years.   Such a massive wind eventually produces a dense circumstellar envelope (CSE)  with density typical  of that observed   in protoplanetary nebulae (PPNe)  and planetary nebulae (PNe) (Herwig, 2005; H\"{o}fner and Olofsson, 2018).
 It is important to specify that these CSEs  are characterized by both relatively weak asymmetries  at a large scale and     a  low velocity field $\sim 10\ km/s$  (Kerschbaum et al, 2010).  However   the observations of  their immediate descendants,  namely the  PPNe   often  exhibit   high velocities associated to a strong bipolarity. This drastic change is     rather intriguing (Duthu et al,  2017). This is the topic of the present paper  to explain how    a very  complex structure, i.e. a structure composed of  an inner disk  or a torus,  surrounded by   a  bipolar inner  structure, itself  surrounded   by  an outer  spherical envelope, can be produced.

\noindent Pascoli (1997) hypothesizes that the primary cause of the ejection of massive winds by an evolved AGB red giant would possibly be the magnetic activity present just above its degenerate core. The transportation of magnetic field from the dense core to the stellar surface is entirely ensured by turbulent diffusivity and, de facto, the modelization was depending on an arbitrary constraint, that is that the macroscopic meridional velocity is zero within the star (self-imposed in a one-dimensional model for which there exists no possiblity to set up a countercurrent flow). A major difficulty with this model is that magnetic stresses have  a smoothing effect on the angular velocity gradient and the dynamo is eventually quenched. In fact, this problem resides  in the zero-meridional circulation hypothesis. However if 
the arbitrary constraint of zero meridional velocity is relaxed, the situation changes drastically. Then a bidimensional
analysis was needed.

\noindent  Nordhaus, Blackman \& Frank  (2007) have succeeded  in developing      much
more  sophisticated  2.5 dimensional models  than the 1-dimensional   model  of 
  Pascoli  (1997).  The  counter-reaction  of the magnetic field  on   the
   differential  rotation is explicitly   considered in these models.  The aim of
these authors was to know if a single star can  produce a  magnetic field  over a
 sufficient   period of time   to eject and to  adequately  shape a PN.  Their
conclusion is  that there exists  a possibility  if a convection reseeds the
angular momentum in an  efficient manner.    These authors   briefly  examine 
this question  in their section 2.4.  Unfortunately no detailed calculations are
presented and no definitive conclusion can be brought.    There exist  indeed 
various types of convections   and some them  can even  destroy the dynamo.   The
only star where a   sustaining--dynamo  convection  is well studied  is the Sun.
  In this star  a  global  convection is existing in the form of  a meridional 
circulation.   This circulation  acts as  a conveyor belt and  is a key     
ingredient in the magnetic activity of the Sun   (Kitchatinov,  2016).  
Following this author :  The meridional  flow  can be defined as a poloidal part
of the global axisymmetric motion resulting from an averaging -- over time or
longitude or ensemble  of convective motions --  of the velocity field.  Without  a 
well adapted self-organized meridional  circulation  the dynamo  cannot  efficiently  run over a long period.
  This mechanism  has a dual   role :  first  to bring  fresh   angular
momentum   toward the region where the dynamo resides and  secondly   to drain
the  newly created magnetic field  for preventing   it   to quench   the dynamo.

{\raggedright
The second   role  is of great importance,    the dynamo cannot  run   if the newly
magnetic field in not quickly  evacuated from  the dynamo zone.   However it has
long been known that  a  magnetic field  cannot  disengage from the star if it  is
diffuse.    For that it must first  be concentrated  (Parker, 1984).  Nordhaus,
Blackman and  Frank  (2007) do not address this  issue.
}

\noindent  In this vein,  Pascoli and Lahoche (2008, 2010) discussed the
conditions of ejection by an evolved  AGB star in the framework of an
axisymmetric magnetohydrodynamic model. A strong toroidal magnetic field is produced by a dynamo mechanism in the core region of the star. The magnetic field effectively tends to brake the rotation in the core region but this effect is now counterbalanced by a supply of new angular momentum from the
higher latitudes  and a steady state can be achieved.  These authors have also shown that the AGB's atmosphere and the ejected circumstellar envelope are not really separable, but should be treated as a continuous mass distribution. 
However a steady state model is not fully predictive in the sense that the solution is very generally ab initio-induced and the equations are simply checked. A much better approach is to take a time-dependent model and  to start from the ending stage of contraction of the core of an AGB star, assuming no circulation and no toroidal magnetic field at the beginning of computations. Then we can analyze how these quantities can spontaneously develop in a self-consistent manner. This issue is explored in the present paper in the framework of time-dependent simulations. The procedure follows the usual appoach of  mean-field electrodynamics (Krause and R\"{a}dler, 1980; R\"{u}diger and Hollerbach, 2004, Charbonneau, 2010). 

\noindent  Another  difference with  the  model of Pascoli and Lahoche  is that the   rotation profile   is now found to have a polar and not an  equatorial structure. Contrarily what we usually think,  it is not so difficult to create a high magnetic field $\sim10^6\  Gs$  in the core region of an evolved AGB star. The problem is  rather   how  to maintain the dynamo  in a steady state  for say here  $10^4   \ years$  (the duration of  the superwind ejection with  a  consecutive circumstellar envelope formation).  If the new field is not rapidly evacuated, it back-reacts on the dynamo and the differential  rotation  is  spreaded  out. The idea is to imagine that the field is built-up  in a region  (the  so-called dynamo region) and the compression of the field takes place in another one. Another problem is linked to the evacuation of the field which must be sufficiently rapid so that   a balance is established between  creation and loss. This time  the field is created  by  a polar dynamo instead of an equatorial one. The  magnetic field is  then strongly compressed in the equatorial  plane. Thus the highly  magnetized  area and the dynamo area become  distinct.  The  novelty here is that now  the action of a  high  magnetic field  does no longer  hinder the dynamo in a natural way. 
 
 \noindent Large-scale magnetic fields in protoplanetary nebulae (PPNe) and planetary nebulae (PNe) were also hypothesized by a lot of authors in order to explain the rather remarkable  morphologies  (Gurzadyan, 1969; Pascoli, 1987, 1992, 1997; Chevalier and Luo, 1994; Matt at al, 2000; Blackman et al, 2001; Garc\'{\i}a-Segura, L\'{o}pez and Franco, 2005). The origin and the effects of magnetic fields in the progenitor were also considered  (Nordhaus, Blackman and Franck, 2007; Busso et al, 2007; Nordhaus et al, 2008;  Blackman 2009; Nucci and Busso, 2014). This paradigm has been considerably strengthened by observational data  of circumstellar envelopes of evolved carbonaceous stars -- the assumed ancestors of PPNe and PNe -- (Vlemmings, van Langevelde and  Diamond,  2005;  Herpin et al, 2006; Sabin, Zijlstra and Greaves, 2007; Kemball,  Diamond and Gonidakis,  2009;  Vlemmings, 2012; L\`{e}bre et al  2014; Sabin, Wade       and L\`{e}bre,  2015; Duthu et al,  2017), but also by laboratory experiments (Ciardi et al, 2009).
 
\noindent   Some authors have also pointed out  the indirect signature of  underlying magnetic fields in PNe by the presence of a large coherent  network of filaments  (Huggins and Manley, 2005). The  matter which composes a filament is naturally oriented following  a directional bundle of magnetic field lines   and  can preserve its coherence over very  long distances (comparable to the diameter of the nebula), and these   filaments are  generally twisted. A typical case is  NGC 3132 where we can very distinctly see a prominent  straight chord composed of two interlacing filaments starting  from a rim and going to the opposite  one of the nebula, see the figure   2 of the paper by Huggins and Manley (is it posssible to produce   similar  structures, i.e. connected on very long distances,  from simple hydrodynamic, shearing  instabilities or shocks ?).

\noindent On the other hand an important point to notice is that if a large-scale magnetic field is existing in the dense  circumstellar envelope produced at the very end  of the  AGB stage (which is  characterized  by a very high  mass loss rate  $\sim 10^{-4}\ M_\odot \ year^{-1}$),  then  this field must necessarily still exist in its descendant, i.e. a protoplanetary nebula (PPN). In other words to observe   a large-scale coherent magnetic field in    dense circumstellar envelopes  of evolved AGB stars implicitely leads to  admit its presence in  PPNe (Duthu et al, 2017).

\noindent We can add that a large scale coherence of the magnetic field in the circumstellar envelopes  signifies  that this field has been created in a very  small region (very likely the core of the evolved AGB star where the rotation is presumably high) and then  extended at  a large scale by  expansion of  the gas   derived from  the star  (it appears  very difficult indeed, if not impossible,  to directly create a  coherent  magnetic field over  a large scale).

\noindent Another very interesting but different  topic to be addressed in the present paper is why the circumstellar envelopes generally appear grossly round at a large scale  (Neri et al 1998; Kerschbaum et al, 2010),  while otherwise  their descendants  (PPNe and evolved PNe)   very often appear strongly bipolar or at least highly   axisymmetric.

\section{The MHD equations}

\noindent We assumed axisymmetry and spherical coordinates (r, $\theta $, $\phi $) are used in all our calculations. 

\noindent The continuity equation is :

  \begin{equation}
     \frac{\partial \rho}{\partial t}+\nabla.(\rho \mathbf{v})=\mathbf{0}  
\end{equation}
    
      \noindent where $\rho$ denotes the mass density and $\mathbf{v}$ the velocity of gas.
  
   \noindent The momentum equation reads:

\vspace{-15pt}
\begin{equation}
\frac{\partial \mathbf{m}}{\partial t}+ \nabla . (\mathbf{mv}-\frac{1}{4\pi}\mathbf{H}\mathbf{H})+(P+\frac{H^2}{8 \pi})\mathbf{1} =\nabla . \mathbf{\pi} -\rho \mathbf{g}
\end{equation}

\noindent In this expression $\mathbf{m}$ is the momentum density ($=\rho \mathbf{v}$), $p$ is the thermal pressure $P$ ($=\frac{\rho k_{B}T}{\mu m_{H}}$ with $k_B$ the Boltzmann constant, $T$ the temperature, $\mu =\frac{1}{2}$ and $m_H$ the atomic unit mass). The isotropic viscous stress tensor $\pi$ which is given by

\[
\mathbf{\pi}=\rho\nu[(\nabla \mathbf{v}+(\nabla\mathbf{v)}
^{T}+\frac{1}{3}\nabla .\mathbf{v}\mathbf{1}]
\]
($\nu$ denotes the turbulent viscous transport coefficient, $\mathbf{1}$ the unit tensor)

\noindent The gravitational acceleration vector $\mathbf{g}$ is taken equal to $-\frac{GM(r)}{r^2}\frac{\mathbf{r}}{r}$. The gravitational mass $M(r)$ of the envelope evaluated at the distance $r$ is calculated with the
relationship:

\begin{equation}
M(r)=M_{c}+4\pi \int_{r_{c}}^{r}\overline{\rho}(r^{\prime })r^{\prime
2}dr^{\prime}
\end{equation}

\noindent where $M_c$ expresses the degenerate core mass and $\overline{\rho}(r)$ designates the averaged-over-latitude density.

\noindent The energy equation may be written in the form (see, e.g., Dobler, Stix, Brandenburg, 2004):

\[
\frac{\partial E}{\partial t}+\nabla . [ (E+P+\frac{H^2}{8\pi})\mathbf{v}-\frac{1}{4\pi}\mathbf{H}(\mathbf{v}.\mathbf{H})]
\]

\vspace{-20pt}
\[
=\nabla .[(\pi.\mathbf{v})-\frac{\eta}{4\pi}(\nabla\wedge\mathbf{H})\wedge\mathbf{H}+ \kappa\nabla T]+\mathbf{m} . \mathbf{g}   \ \  \  (3)
\]

\noindent where $T$ denotes the temperature,  $\kappa $ represents the thermal conduction coefficient. 

\noindent Eventually the kinematic axisymmetric dynamo equation is (Dikpati and Charbonneau,1999; Charbonneau, 2010):

\vspace{-15pt}
\begin{equation}
\frac{\partial \mathbf{H}}{\partial t}+\nabla \wedge (-\mathbf{v} \wedge \mathbf{H})= -\nabla \wedge  (\eta \nabla \wedge \mathbf{H}- \alpha \mathbf{H})
\end{equation}

\noindent where $\eta$ is the turbulent magnetic diffusivity ($\eta =\frac{1}{3}\tau \overline{\mathbf{u}^2}$ with $\tau$ the correlation time for the turbulence and $\mathbf{u}$ the turbulent velocity).

\noindent The source term on the right hand side $\alpha \mathbf{H}$ ($\alpha =-\frac{1}{3}\tau \overline{\mathbf{u}.\nabla\wedge\mathbf{u}}$) expresses the regeneration of the field by isotropic $\alpha-$mechanism (See, Dikpati and Charbonneau, 1999; Charbonneau, 2010).

\section{computional details}

\noindent All the calculations were performed using the PLUTO package (Mignone et al, 2007; Mignone et al, 2015) implemented on a SGI Altix UV100 computer. This program is a finite-volume / finite-difference, shock-capturing code designed to integrate a system of MHD-equations in the form of conservative laws (the MHD equations given above are discretized using  this form). Flux computation has been made employing the hhlc (Harten, Lax, Van Leer) Riemann solver. The RK3 time-marching algorithm was used. The domain under study is extended
from $10^{9}$ cm to $3\ 10^{15}$ cm.  This domain is thus divided in three subdomains :

\noindent a. the core (dynamo) region from $r_{c}=10^{9}$ $cm$ to $3\ 10^{12}$ $cm$. For this area, the various time-scales at stake are very different: the dynamical   time scale  $\sim  3  \ s$, the rotational period  $\sim 600 \ s$,  the characteristic time of the meridional circulation $\sim 10^3 \ s$ and eventually  the characteristic times of diffusion  and  regeneration of the poloidal field from the azimuthal component, both   $\sim 10^5 \ s$.

\noindent b. the transition area from $3\ 10^{12}$ cm to $r_{\star}=3\  10^{13}$ $cm$. 

\noindent c. the circumstellar envelope from 
$3\ 10^{13}$ $cm$ to $r_{\infty }=3\ 10^{15}$ cm.

\noindent These three domains  form a single one,  but this arbitrary division is imposed by the characteristic time  of evolution  which is once again different for  each of these regions. It is thus important to notice that such a    division  is a  practical way to greatly shorten   the   CPU time (by reducing it to only  a few  months !). This division results from a numerical treatment  but does not necessarily express    a physical reality.  An ideal situation would  obviously be to treat these three zones   as an unique one, but the latter  procedure  would need  a very big supercomputer.

\noindent The parameters :

\noindent We have assumed that the turbulence (velocity $\mathrm{v}_{t}$) is exalted
in the vicinity of the core (radius $r_{c}$):

\begin{equation}
\mathrm{v}_{t}=\mathrm{v}_{tc}(\frac{r_c}{r})^{1/2}+\mathrm{v}_{te}
\end{equation}

\noindent Exaltation of the turbulence is predictable taken account of the impinging stream of matter at the base of the convective envelope, this stream resulting from the descending polar column linked to the contraction of the inner region at a final stage.

\noindent The parameter $\mathrm{v}_{tc}$ ($=$ $5\ 10^{5}$ $cm$ $s^{-1}$) represents
the turbulent velocity taken at the core surface and $\mathrm{v}_{te}$ ($=10^{5}\ cm$ $s^{-1}$) is the turbulent velocity throughout the 
evolved AGB's envelope.

\noindent The $\alpha -$effect (restricted to the isotropic case) is modeled by the closed-form function:

\begin{equation}
\alpha =\alpha _{c}\frac{1}{
1+(\frac{H}{H_{eq}})^{2}}(\frac{r_c}{r})^{1/2} cos\theta
\end{equation}
with $\alpha _{c}$ $=10^{4}$ $cm$ $s^{-1}$ and the equipartition value $
H_{eq}=\sqrt{4\pi \rho }$ $\mathrm{v}_{t}$. This effect is responsible for
the production of magnetic fields by $\alpha$ process as admitted in the
framework of the mean-field dynamo theory. The quenching factor $1+(\frac{H}{%
H_{eq}})^{2}$ in the denominator underlies that the Lorentz force associated
with the dynamo-generated magnetic fields impedes the turbulent fluid
motions. This term ensures that the poloidal field generation process is
stopped when the toroidal component is close to -or higher than- the
equipartition value $H_{eq}$ (See, e.g., K\"{u}ker, R\"{u}diger and Schultz,
2001). The other factors, that is the $cos\theta$ latitude profile and the
exponential r-dependence, are of common use in the mean-field dynamo models
where they appear as simple geometric cutoffs (See, for instance,
Chatterjee, Nandy and Choudhuri, 2004; Charbonneau, 2010).

\noindent Similarly, we define the profile of $\eta$ by the expression:

\vspace{-15pt}
\begin{equation}
\eta =\eta _{c}\frac{r}{r_{c}}\frac{\mathrm{v}_{t}}{\mathrm{v}_{tc}}\frac{1}{
1+(\frac{H}{H_{eq}})^{2}}
\end{equation}

\noindent where $\eta _{c}=10^{13}$ $cm^{2}s^{-1}$ (magnetic Reynolds number $R_m$ $\sim$ $100$).

\noindent As mentioned above, the term $\frac{1}{1+(\frac{H}{H_{eq}})^{2}}$ represents
the $\eta $-quenching as due to the nonlinear magnetic field feedback on the
turbulence. The H-dependence of $\eta $ introduces a dose of nonlinearity in
the problem. A quite similar factor is used in the solar dynamo theory
(Dikpati and Charbonneau, 1999; Gilman and Rempel, 2005; Charbonneau, 2010). The $\frac{r}{r_{c}
}$ factor traduces the reasonable hypothesis that the mixing length linearly
increases with $r$ in the evolvd AGB's envelope (Pascoli, 1997). The present calculations thus fully assume that the MHD physics available for the Sun is
immediately transposable to evolved AGB's stars. It is difficult to say if it is
effectively the case, but it seems that this physics appears universal and applies to all structures in the Universe, whatever the characteristic length of the object under examination (more specifically all types of stars) is. These considerations are general and have also been developed outside the strict solar domain (Beck et al, 1996).

\noindent We admit that the magnetic Prandtl number $P_{r}=\frac{\nu_c}{\eta_c}=1$. All these parameters have dimensions of $L\times V$, or numerically can be estimated  $\sim 0.3 l_c v_{tc}$ with $l_c$ turbulent correlation length) (Pascoli and Lahoche, 2008, 2010). However other choices can still be made (see for instance Nucci and Busso, 2014).

\noindent  The pertinence of the values chosen for the adjustable parameters $A_c$ and ($\eta _{c}$, $\alpha_c$)  must lead  to agreement with the observational data, especially here the measurable quantities, i.e. mass loss, expansion velocity and flux loss produced by the evolved AGB stars during the slow superwind phase. This statement does not vindicate the model in itself but at least shows that, within its explanatory framework and with realistic numerical values for these parameters, we can obtain the good values for the observable quantities (As in other similar MHD problems, for instance in the dynamo models for the Sun, where the aim is to obtain the best fit to observed solar cycle).
 
\noindent In order to solve the system of equations, the initial conditions have still to be specified. We have chosen a simplified model of evolved AGB star. 

\noindent The total mass of the star $M_{\star}=M_{c}+M_{env}$ is taken equal to
one solar mass, with a core mass equal to $0.5$ $M_{\odot}$ ($r_{c}=10^{9}$
$cm$) and a mass for the convective envelope equal to $0.5$ $M_{\odot}$ ($r_{\star}=3\ 
10^{13}$ $cm$). 

\noindent The initial density  and temperature in the evolved AGB's envelope is taken by solving the equilibrium equation (assuming both spherical symmetry and no macroscopic motion in the convective envelope) :

\begin{equation}
-\frac{GM(r)\rho(r)}{r^2}-\frac{\partial P}{\partial r}=0
\end{equation}

\noindent We have assumed that  the matter  composing   the convective envelope   is essentially an ideal gas with   negligible radiation pressure.  This approximation appears legitimate  at an average  point of the envelope, even though this one becomes   false  just  above the CO core (more  precisely the base of the convective envelope). In spite of this we have put   $P\equiv P_{th}$  everywhere  in the convective envelope (after noticing that, approximately,     $P_{th}\sim r^{-\frac{5}{2}} \gg P_{rad}\sim r^{-4}$).   This  trick  (associated to many other drastic simplifications made in this work !) helps to save CPU time which still remains  very long. In fact the essential problem is numerical.  The consideration of the radiation pressure   creates  a much higher gradient  of density  above the core surface and the numerical mesh has to be considerably reduced (increasing the CPU time).

\noindent We admit that  this subterfuge  should not  affect the dynamo too much. On the contrary we may  think  that  the consideration   of the radiation  pressure would  allow  to increase  the density (and the pressure) above the  CO core position (compared to the values taken  in our model)  and thus to  still minimize  the harmful back-reaction on the  differential rotation (the source of the field).  

\noindent That being said,  the pressure-density-temperature relation is here  the usual ideal law $P=2\frac{\rho}{m_H} k_B T$, available for a fully ionized medium composed of pure hydrogen. The density and temperature  at $r_{c}=10^{9}\  cm$  are respectively $\rho_c=0.1 \ g  cm^{-3}$ and  $T_c= 1.5\ 10^8\ K$.

\noindent  We must also specify  that what is considered here as  the base of the convective envelope   is not directly   identified  with the CO core surface. The region just above the inert CO core (radius $r_c = 10^9\ cm$)  is  indeed the seat of  very  complex  nucleosynthesis processes.  We find there a thin double sheet  consisting in a He-burning shell surrounded by a H-burning shell.  These two thin sheets, which can alternatively be active or not,  are separated by an intershell consisting of a mixture   of  $He$, $C$  with  a few percent of $^{22}Ne$ and  $O$ (Busso, Gallino and Wasszeburg, 1999, Herwig, 2005; Karakas and Lattanzio, 2014).   What is      named here "base of the convective envelope"  is located just above this  double sheet. Let us note that the inner zone, i.e. the inert CO core and the He and H burning shells  on top of it,   do not intervene   in the specific scenario  envisioned  here,  that is the build-up  of a toroidal magnetic field at the base  of the convective envelope, its transport through this envelope and eventually  its ejection  accompanied by  a strong mass loss at the AGB surface. If, possibly, there exists     a   core dynamo, this one  is then assumed to be fully   disassociated from   that   described in the present paper.  The magnetic field  considered here is not anchored on the CO core surface. This will be discussed further below.

\noindent  At the "stellar surface" the  boundary  conditions are  $M(r=r_{\star})=M_{\odot}$, $\rho(r=r_{\star})\sim \ 10^{-11} g \ cm^{-3}$ and $T(r=r_{\star})=3000\ K$. These values are slightly different from those chosen  for instance by Nucci and Busso (2014), but it must be remembered that they are arbitrary  (the surface of an evolved AGB star is not a geometric surface perfectly defined. It is rather a "volume" with an thickness  of  $\sim 3 \ 10^{12} \ cm$. For the region of emission of the slow superwind, there is not even any surface at all, but a continuum between the "star" and an outer thick disk).

\section{Results}

The output data (with format .vtk) have been visualized using the open source package VisIt (version 2.10) distributed by the Lawrence Livemore National Laboratory.

\noindent a. The core region

\noindent We name   here by   "core region"  the zone  located just   above the system composed of   the inert   CO  core crowned by  the H-He double sheet. It is in fact  the deepest   part of the  convective envelope.   The rotation profile  within this   region is an  essential  data for the dynamo models. This is indeeed  the differential rotation which insures the build-up of the magnetic field in this region. Retrospectively  the present model could thus  help to fix the initial conditions for the rotation and the magnetic field in the core region of an  evolved AGB star.

\noindent  Fortunately even though the Sun and an evolved AGB star  are two  very different stars,  the Sun will evolve in an evolved  AGB and it appear reasonable to assume  that some imprint  seen  in the rotation profile of the Sun will still  exist    at the  evolved AGB stage  (not least by angular momentum conservation).    The Heliosismology produces the internal differential rotation profile of the Sun.   The  physical conditions reigning   at the surface of the Sun are equally  known.      In the Sun the  angular rotation is quasi uniform  with a value  at the surface of the order of $2\  10^{-6}  s^{-1}$.   We take at $r=10^{11} \ cm$   $\Omega_{11}\sim 2\  10^{-6}\  s^{-1}$.   This numerical  value appears reasonable.  Concerning the solar core region   recent measurements of  the  rotation rate by identification of asymptotic gravity modes (GOLF instrument on board SOHO)  (Fossat et al, 2017) ) seem to contradict the  preceding values  obtained by BISON  (Paterno, Sofia and di Mauro  1996).  If  the results  of Fossat et al  are confirmed  the angular velocity of the  solar core would be  much higher than at the surface (by a factor $4$), increasing at the same time the available total angular momentum. Obviously the value observed at the surface is left unchanged.

\noindent We  do a similar reasoning  for the magnetic field.  For the Sun the  observed magnetic  flux  is of the order of   $10^{23}\  Mx$.  For a   radius of the order of   $10^{11}  cm$  this gives a mean magnetic field of $1$ Gauss. A   series of preliminary runs shown that this initial  magnetic field concentrates  in the  polar area  with conservation of flux. This eventually gives   an estimate of  $\sim 10\ Gs$ for the magnetic field for the latitudinal angle  $\theta=5-30^{\circ}$.  We have then started our calculations with a guess entry for the  magnetic field intensity imposing a  value of $10$ Gauss  at  say  $r_{11}= 10^{11}\  cm$ for the latitudinal angle $\theta$ lying  between $5-30^{\circ}$. 

\noindent In our preceding work (Pascoli and Lahoche, 2010) the model  was steady  and we taked   the boundary  conditions (magnetic field and rotation)   at the surface of the  core.  Instead, here,  the model  is    time-dependent  and the boundary   conditions  (magnetic field and rotational velocity)   are taken at   $r_{11}=10^{11} \   cm$. This procedure  now appears  much more natural because  this is the sense of  the meridional circulation  which carries the angular momentum from the  outer polar regions toward the base  of the convective envelope (the radius of this base is labelled $r_c$ in the following). 
The distribution of angular velocity is thus modelized from that of the Sun  after contraction  of the central region, i.e. with respect to  the analytic formula $\Omega(r,\theta)=\Omega_c (\frac{r_c}{r})^2$ (obeying to  the law of angular momentum conservation with variation in $r^{-2}$ during the contraction of the core).  In this context  the angular velocity at the surface of the evolved AGB star is very weak $\sim 10^{-9} s^{-1}$ or  $v_{eq} \sim 0.3$ $km s^{-1}$ (equatorial value for the azimuthal velocity) , while near  the base of the convective envelope  $\Omega_c\sim 10^{-2}$ $s^{-1}$ or $v_{eq} \sim 100$ $km s^{-1}$. This quantity is a key parameter of the model.  We can remark that the  mean equatorial velocity of the white dwarfs is sensibly weaker  $\lesssim 20\ km s^{-1}$.  We thus implicitely assume     that the region surrounding  the core of an evolved AGB star rotates more rapidly than a white dwarf  and/or still that it is this core itself which slows down during the envelope ejection (by magnetic coupling but a very high magnetic field is needed). May be  can we  eventually imagine more simply that the degenerate core rotates more slowly that the base of the evolved AGB's envelope, where the magnetic field under consideration here is generated. We can suggest a vague analogy with the superrotation of the Venus atmosphere which rotates much speeder than  the  planet itself, even though in this case  the latter phenomena appear  in another very different context  and   have  of course a distinct  origin (planetary polar vortices has also been found in the solar system, however  the mechanism which maintains them  seems a delicate balance between warming and cooling in the atmosphere. On the other hand as due to very low conductivity no  dynamo (and no magnetic field) is associated to these vortices).

\noindent As for the magnetic field we start with a mixed extended dipole-quadrupole poloidal component represented by the potential vector (a seed of field is needed to get the dynamo started) :

\vspace{-10pt}
\begin{equation}
A_{\varphi}(r, \theta)=A_{11}[1-exp-e^{-(\frac{r-r_{c}}{
0.1r_{c}})}][0.1(\frac{r_{11}}{r+r_{11}}) sin\theta
\end{equation}

\vspace{-15pt}

\[+0.9(\frac{r_{11}}{r+r_{11}})^3 sin\theta cos\theta]
\]
The boundary conditions at $r\sim 10^{11} cm$ are $H_r   =-10\  Gs$ for $\theta=5-30^{\circ}$ and $=0$ for $\theta=0^\circ-5^\circ$ and $30-90^{\circ}$ (and the same thing in the south hemisphere by symmetry but with $H_r  =-5\  Gs$).  The constant $A_{11}\sim 6 \ 10^{12} \ Gs \ m$ is estimated from the  value at  $r\sim 10^{11} cm$. 
 
 \noindent  Let us notice that the magnetic field value measured by Jordan, Werner and O'Toole (2005) at the surface of  a small number of central stars of PNe (the evolved remnant of the evolved AGB's  core), that is of the order of $10^3$ $gauss$,  does not intervene in this initial conditions (we avoid to fix the boundary conditions for the magnetic field at $r=r_c$ which are unknown). The values of Jordan et al are measured at the surface of the central star of PNe, not at the surface of the core of an evolved AGB star (where this measurement is definitely not possible). If there exists a magnetic field built-up by the degenerate core, it  is very likely disconnected from this one which is    considered here and   which is  produced  at the base of the evolved AGB's envelope. In any way and after new observations  the intensity of the magnetic field measured at the surface of  central stars (the bare core)  of  PNe   seems  weak ($\lesssim \ 1\ kGs$, see  Jordan et al, 2012)).  Paradoxically enough some white dwarfs ($\sim 10$ percent)  have a surface field  ranging from $10^3$ up  to $10^9 \ Gs$. However  for high magnetic values,  field generation within the common envelope of a binary stellar system has been suggested (with production of a complex and non-dipolar structure exhibiting the presence of higher order multipoles) (Ferrario et al, 2015).

\noindent Concerning the dipole component, this one  plays a minor role and could eventually be suppressed. This component  forms two sheets of opposite signs in the equatorial  plane  where finally   it annihilates. Besides it creates an north-south asymmetry between the two hemispheres.  Sole the quadrupolar component contributes to the formation of the magnetized equatorial disk. In any way the initial conditions taken for  the field are not very  significant because this field  is strongly reshuffled  by the meridional circulation toward the pole.

\noindent   We have started  the calculations with the initial conditions  supplied above. No meridional circulation and no toroidal magnetic component  is assumed to be present.

\noindent Especially important is the root cause  of  a  one-cell     meridional circulation which  appears  in the convective envelope (we know that a similar phenomenon is existing   in  the Sun and which extends from the   tachocline up to the surface, see   Liang et al, 2018).  In the present model, we are mainly concerned by the apparition  of such a  circulation  but    at the tip of the AGB phase (the stage where the  AGB  becomes  a so-called   OH/IR  star).   Very likely  this self-organized   motion starts in a zone located  above the  CO core -- double H-He burning  shell system and then extends  over   the totality of the convective envelope.   The causes of a locking of chaotic small-scale motions in  an organized mode at  a large scale can be numerous (Kitchatinov, 2016). We can indeed  invoke  an  ultimate slight core contraction which can reinforce  the rotation,  a  gradient of temperature  or still a gradient  of chemical abundance  between the equator and the pole. In fact very small effects can produce  a large-scale circulation. In the following we  favour the  simple process of a slight core contraction (with  a consecutive increase of the rotation in the core region)  for the  initiation     of the meridional circulation.  The driving of a meridional flow by centrifugal forces is long known (Kitchatinov, 2016).     Let us note that  in the Sun  the  apparition and the   persistence    of the meridional circulation in the convective zone  is also a  problematic area.  In the MHD models  for the Sun  the meridional  circulation is  directed    by the  inconspicuous  introduction at the beginning of the calculations   of  a lot of  fine-tuned  parameters, analytic formulae, etc   (e.g. Choudhuri,  Schussler, Dipkati, 1995;    Charbonneau,  2010;  Kitchatinov, 2016). The same procedure is used in in other stars   (Dobler, Stix, Brandenburg,  2005). However, very interestingly,   we will see  later that,   once   formed,  this meridional circulation coupled to the differential rotation in turn triggers   the production of a strong toroidal magnetic field which ultimately leads to a  very high mass loss by the star.

\noindent We assume thus here that the final stage of contraction of the central regions triggers  an amplification of the angular velocity above  the inert CO  core.  Our numerical simulations show that this phenomenon  is  spontaneously accompanied in its turn by the formation of vortices with strong mean velocities ($\sim 5\   10^5-10^6$ $cm \ s^{-1}$). A surimposed large scale meridional circulation also appears but with low mean velocities ($\sim  10^5$ $cm \  s^{-1}$ at $r=10^{11} cm$). We can notice that this large scale flow is not here assumed by using an adhoc streamline function like in most solar models (Guerrero, Mu\~{n}oz and de Gouveia Dal Pino, 2005; Charbonneau, 2010;  Pipin and Kosovichev, 2011;  Nucci and Busso, 2014), but its directly results  from computations (the initial conditions at $t=0$ assume no meridional circulation). Even though the procedure is extremely time-consuming, here the velocity field is not chosen in advance and the model  is  self-consistent. We note that the meridional velocities calculated here $\sim  10^5\ cm \  s^{-1}$ at $r=10^{11} cm$ are much higher than that measured  at the surface of the Sun, $\sim 10^3   \ cms^{-1}$. However the centrifugal excitation is also much higher here and is appears natural that the meridional circulation be exalted.  A single cell is  generated by hemishere  (this statement seems also favored in the convective area of the Sun, as suggested by SOHO/MIDI and SDO/HMI observations, but there exists other possibilities (see Liang et al, 2018), even though such an analogy  must obviously be taken with cautious being  the Sun and an evolved AGB star  are very different in their structure).

\noindent At the beginning  of calculations ($t\sim 0.25 \ h$ is  of the order of  the period of rotation of the degenerate core) the dynamo is both  equatorial  and polar (the figure  2.1 shows the beginning of the magnetized disk formation). However the equatorial dynamo produces a  weak azimuthal transient  field with an  opposite polarity, which rapidly disappears while  the polar contribution goes down to the equator and becomes  rapidly dominant by compression.

\noindent After a duration $\gtrsim \ 1 \ h$, examination of fig. 1.2  shows that the initial angular rotation has been deeply reshuffled by the meridional circulation. It is now largely "antisolar", i.e. very fast at higher latitudes (the pole region), but much   slower  near the equator  as due to the strong magnetic breaking (fig. 1.3). Modifying by a factor $2$ or $1/2$  the key parameter of the model, that is the initial angular velocity at $r=10^{11}\ cm$, $\Omega_{11}$, does not change this situation.

\noindent  Quasi-simultaneously, the toroidal magnetic field component is generated from the poloidal field at higher latitudes ($\gtrsim 60^{\circ}$) as due to shearing by axisymmetric differential rotation or $\omega$-effect (figs. 2.2, 2.3). The meridional vortices have then a rolling mill action which squeezes and amplifies, with constant magnetic flux,  the  magnetic field toward the equatorial plane where it is finally stored forming a strongly magnetized disk. We can thus notice that the process protects the differential rotation which is located at higher  latitudes against spreading by the Lorentz forces, the dynamo region and the storage area being distinct.  The built-up of toroidal vortices and magnetic buoyancy are not  interlinked (we recall that the toroidal vortices are already present before the built-up of a significant magnetic field, in other words the meridional circulation appears first and then after the azimuthal component of the magnetic field is created. This result was not accessible in the preceding model of Pascoli and Lahoche where the dependence in time was not taken into account).  In addition, the equatorially-concentrated flux loss from the dynamo region is the dominating field-limiting process for the dynamo (by contrast, the magnetic diffusivity in the polar column has just a smoothing role). Likewise the gas in the thin magnetized disk is dragged by buoyancy and flux ropes present in this disk float upwards. 

\noindent After a time  of the order of a few     hours      the initial  configuration of the magnetic field has thus been   deeply transformed by the meridional  circulation.  The  rotation is now essentially polar and the poloidal magnetic field  is  also largely  concentrated   in a polar column  within an angle  $≤30^\circ$ (fig. 3).  Let us note that the boundary condition  $v_r=0$ at  $r=r_c$,  $\theta=0^\circ$ (the core is  assumed to be impenetrable)   imposes that a static column  is necessarily present above the core within an angle  $\lesssim 5^\circ$ with no field.   In the polar  column the  magnetic field is  azimuthal  just above the core and quasi-radial at a large distance (fig. 4).

\begin{center}
\includegraphics[height=160pt,width=170pt]{mag_field-eps-converted-to.pdf}
 \end{center}
 
\noindent Figure  4  : Sketch   of an  idealized dynamo-generated  magnetic flux  above the core region inside an evolved AGB star

\noindent Once installed the  rotation field   does not longer vary because the magnetic torque remains weak.  The equilibrium value for this torque  is   (near the core)

\vspace{-5pt}
\begin{equation}
{(H_r H_\varphi)}_{eq}\sim 4\pi(\rho v_r v_\varphi)_c
\end{equation}

\noindent or  evaluating  the  right  member  $\sim 10^{14} \gg (H_r H_\varphi)_c \sim 10^{11}$.  Fitting  laws for the physical quantities\footnote{Only the  the numerical coefficients result from the fitting, the litteral terms stem from dimensional analysis.}  in the polar dynamo column for $r>1.2\ r_c$ and $5^\circ\leq \theta \leq 30^\circ$  are (the flow is quasi-radial  and the poloidal magnetic field is practically reduced  to its radial component):

\vspace{-10pt}
\begin{equation}
v_r \sim r^{-1/2} \  \  \  \Omega=\Omega_c [(\frac{r_c}{r})^2-\alpha(\frac{r_c}{r})^{7/2}]  
\end{equation}   

\noindent     with $\alpha=\frac{(H_r H_\varphi)_{c}}{(4\pi(\rho v_r v_\varphi)_c)} \sim 10^{-1}$ and

\begin{equation}
 H_r \sim r^{-2} \  \    \  \   H_\varphi \sim r^{-5/2}
\end{equation}

\noindent and

\vspace{-20pt}
\begin{equation}
 \rho \sim r^{-3/2} \  \    \  \   T \simeq T_c[(1+\beta) \frac{r_c}{r}+\gamma (\frac{r_c}{r})^{7/2}]
\end{equation}

\noindent     with $\beta=-\frac{1}{10}\frac{mv_c^2}{kT_c}\sim 10^{-5}$ and $\gamma=-\frac{3}{10}\frac{m H_{\varphi c}^2}{4\pi\rho_c kT_c} \sim 2\ 10^{-6}$. We can note that the descending flow from the pole results in   a very small lowering  of the temperature above  the core. These laws are approximately  accurate for  $r>1.2\ r_c$ and $5^\circ\leq \theta \leq 30^\circ$.

\noindent  Even though  not exactly in the same context,  Nucci and Busso (2014, fig. 2 and eq. 17) grossly suggests  similar simple  scaling laws for the evolved AGB star's convective envelope (without the  correcting  terms proper to the dynamo region which is not considered by these authors). However there exists  an important difference for  the circulation which decreases as a  function of $r$ in our  model  but is increasing   in their work.
 
\noindent Starting from these laws, a semi-analytical model is certainly possible at least for the polar vortex (the dynamo area). The  great interest of a semi-analytical model would be to skirt     the numerical simulations which are hugely  time-consuming.

\noindent Let us notice that the rotational law  is perturbed but just   a slightly  by the presence of the magnetic field (compared to the initial simple law in ${r}^{-2}$). The decrease of the gradient of differential rotation due to the presence of the magnetic field is low. A steady polar dynamo  thus takes place  and this one  is expected to last  for a period of at least $5\ 10^3\  years$ (a very short period before  the totality of the AGB phase and what could be constitute the final stage with a massive expulsion of matter). This period corresponds to the ejection of the totality of the convective envelope with mass loss $\sim 10^{-4}\  M_\odot\ year^{-1}$. A shutdown of the meridional circulation would obviously lead to a cut off  of the dynamo. If the first ingredient for the dynamo  is the presence of a polar vortex, a stable meridional circulation is the central piece  to sustain  this  dynamo. This circulation has  also  to be in the right sense (the polar vortex   is constantly fed with  fresh  angular momentum by the meridional  circulation).  Seen from the core surface  the matter rises at the equator and sinks near   the pole  axis.  Without these two essential conditions  (a polar vortex and a meridional velocity taken in the right sense) the dynamo does not run. On the other hand it is very difficult to know if such a phenomenon can  last for  $10^{4}$ years. We note  that the presence of a meridional circulation is also attested in the Sun  (where besides the same intricate problems to solve  are encountered).

\noindent The magnetic field is maintained at such a moderate  level ($H_{\varphi_c} \sim 10^6\   Gs$)  owing to  the high    flux loss   along  the equatorial plane which evacuates the field\footnote{Let us specify that the term of "flow" of magnetic flux  would be more adequate because the term "loss" seems  to suggest that   the magnetic field is destroyed. It is not the case. The flow of magnetic flux is conservative. It is well known for a long time that the turbulent diffusion is unable  of significant destruction of magnetic fields in a star (the turbulence is quenched by the magnetic field  at a small scale well above the molecular level where destruction by ohmic diffusion is indeed  possible, but this one  is negligible here), see for instance Vainshtein and Rosner, 1991. Eventually  it is experimentally proven that turbulence can even reduce the magnetic diffusivity and concentrates the field (Cabanes, Schaeffer, Nataf, 2014).}.  Clearly  the field is rapidly created by a polar  dynamo and  the equatorial evacuation  insures a  steady state (for at least a period of $10^4\ years$  which is indeed a short period before the  total  duration of the AGB stage).  Jordan et al (2012) have shown that the magnetic field of central stars of PNe is weak. In fact when the evolved AGB envelope is ejected, the dynamo mechanism described here  simply disappears together with the envelope. A refined  analysis of our  results near the surface of the degenerate core shows that  the  azimuthal magnetic  field lines    slide on  this   surface  and  that the  poloidal component  is not anchored to  it.  The magnetic field of the core could even  be  null. The idea  that the magnetic field of the central stars of circumstellar envelopes  is  linked to that of  their  surrounding  circumstellar envelopes must be  give up.

\noindent Near the core surface  the magnetic field is quasi azimuthal $\sim 10^6  \ Gs$. The temperature slightly increases and the density decreases  toward the equator following the fitting laws

\vspace{-20pt}
\begin{equation}
 \rho \simeq \rho_c(1+\delta sin^2\theta) \  \    \  \   T \simeq T_c(1+\beta + \gamma)(1+\epsilon sin^2\theta)
\end{equation}

\noindent     with $\delta =-\frac{mv_c^2}{2kT_c}\sim 4\ 10^{-5}$ and $\epsilon=9\frac{mv_c^2}{2kT_c} \sim 4\ 10^{-4}$. 

\noindent  The magnetic field is also quasi-azimuthal in the equatorial plane. It is strongly concentrated in a  thin    disk (with  a  noticeable uniformity guided by the meridional rollers  which squeezed  the disk)   where  the diffusivity is very low.  The density deviation within the disk   with respect to the field-free   surroundings  is weak   $\delta\rho/\rho \sim 4 \ 10^{-5)}$,  but this effect  however is sufficient to insure  the natural  uplift of the matter opposing  a positive force  to the retraction force  of the magnetic field lines.

\noindent We can also note that the energy distributed  in the meridional circulation $\rho v^2 \sim  10^{11}\  erg\  cm^{-3} $  and in the  magnetic fields $H^2/{8\pi}  \sim  10^{11}\  erg\  cm^{-3} $ in the core region are weak  before the gravitational energy  $GM_c\rho/r_c \sim  \ 10^{16}\  erg\  cm^{-3} $   (magnetic intensity equivalent $\sim  7\ 10^8    \ Gs$).   Both the meridional  circulation and  magnetic  fields only  very  little  affect     the core region   equilibrium. 

\noindent The density distribution remains spherical  in the convective envelope excepting in the region where a strong azimuthal magnetic field is present (but the deviation is $\frac{\delta\ \rho}{\rho}$ is very weak  $\sim \ 4  \ 10^{-5}$).

\noindent The disk  remains thin up to  $3\  10^{12} \   cm$.  Its thickness slowly  increases as  $r^{0.58}$  and  otherwise the field decreases as  $r^{-0.66}$.  The transport from the core  up to this distance  takes $\sim 25 \      d$. During this transport  the orthoradial balance between the disk and  its surroundings is  checked. At the same time, the ratio of magnetic energy to thermal energy within the disk drastically increases from a very weak value  $\sim 10^{-5}$ to $\sim 1$. The fact that the field is not diffuse  but  remains   concentrated  during the transport   is a  very happy situation  because a diffuse (and  therefore weak by flux conservativity) field would not supply an  efficient     mechanism  for the ejection of gas  and, furthermore,     it appears especially  difficult to  concentrate  a diffuse  field at the surface.  Very similarly the observations  at  the surface of the   Sun show  that the solar  magnetic field appears    to occur not in  a  diffuse  configuration  but rather in  small-scale concentrations of  very high   intensity.

\noindent  On the other hand  while the disk is quasiuniform, at both of  its boundary  surfaces  the magnetic field is weak and can be manipulated by turbulence at small scale with bubbles emerging  in the free-field medium. Subsequently  the uplift of the disk must be  accompanied by  an orthoradial mass loss (may be following  a process similar to  the loss of plasma by the coronal holes at the surface of the Sun (see Parker, 1984)).  Unfortunately     this lateral mass  flux is very   difficult to quantify and this is left as a free parameter. Let us specify that this phenomenon is not accompanied by  a concomittant net  lateral magnetic flux loss because the magnetic field lines in the bubbles which randomly  rotate become  diversely oriented and mutually annihilate (see for instance Vainshtein and Rosner, 1991, especially their figure 1). Very likely other mechanisms which  concentrate the magnetic field during the transport up to the surface  are possible (Brandenburg, Kleeorin and Rogachevskii, 2016).   The radial flow of  magnetic flux  in the disk is conservative (i.e. $\dot{\Phi}$  is constant). If one relaxes the axisymmetry hypothesis, that is taking into account of the $\varphi$-dependence, the disk could undulate during the transport but the analysis of such a phenomenon requires to work in the framework of a 3D model and a supercomputer is needed.

\noindent A  very compelling  result of this model  is that  starting with solar values  at the beginning of the calculations,  and after  of  while,  waiting up to the installation  of  a steady state,  we obtain  a  flux loss $\phi\sim 10^{20}\   Mx\  s^{-1}$  which is typical of magnetic flux observed in circumstellar envelopes (for a circumstellar envelope with a mean  radius  of $10^{16}-10^{17}  \ cm$  and a mean velocity of  $10\ km\  s^{-1}$ a mean magnetic field of $10^{-3}\  Gs$ (observational data)   the flux is  $\phi=10^{19} - 10^{20}\   Mx\  s^{-1}$. Is it a coincidence ?).  This statement  strongly supports    the proposed   model : it is then tempting to assume that a single star with a  solar  mass   is    able to   produce at the tip of the  AGB phase the magnetic  field  flux  observed in  circumstellar envelopes of evolved stars. We can note that this machinery  can also run with a wide binary star but may be with interesting addition of  precessional effects  very often invoked by observers of PNe  (Balick et al, 2001).

\noindent b. The evolved AGB envelope-superwind transition area

\noindent  We have introduced a so-called  "transition zone"  which begins at $r=3\  10^{12} \ cm$. This terminology, has been  already considered  earlier in the text as a mesh division,  but also calls for a physical explanation. This zone is part of the convective envelope and is defined as follows.  Near the core surface the ratio of the  magnetic pressure, $P_{mag}$, to thermal pressure,  $P_{th  }$, is of the order of  $10^{-5}$. The  magnetic field plays a little role in the vicinity of the core surface.  In the inner part of the  convective envelope the thermal pressure approximately  decreases as  $r^{-\frac{5}{2}}$ while the magnetic pressure  decreases more slowly  as $r^{-1.3}$ (the magnetic field is squeezed in the equatorial plane by the rolling mills action of the meridional circulation). We arbitrarily fix the base of the "transition zone" where $\frac{P_{mag}}{P_{th}}\sim 10\  percent$.  When approaching the surface of the star this  ratio still continues to increase  and  the magnetic pressure becomes comparable to the thermal pressure, which ultimately leads to an MHD ejection.

\noindent  The "surface" of the evolved AGB star is not a so well defined simple geometric shape.  Rather it is a "volume"  which  extends from $3\ 10^{12}\ cm$ to $3\ 10^{13} \ cm$. The boundary  conditions at $r=3\ 10^{12} \ cm$ are $\rho_e=5.3 \ 10^{-7} \  g \ cm^{-3}$, $H_e=0$  in the evolved AGB's envelope and    $\rho_i$  is left undetermined (for $\rho_i=\rho_e/2$, a mass loss much  higher than  $\sim 10^{-4} \ M_\odot\ year^{-1}$ could be  reached),  $H_i=5   \  10^3\   Gs$ (by conservativity of magnetic flux) in the thin disk.  Only a narrow  equatorial band  is concerned with angles  between $-15^\circ$ and $15^\circ$. For both these sides  the conditions of the static envelope have  been  prescribed.  A reflective condition is assigned  at $\theta=0$. At $r=3  \  10^{13} \ cm$  the boundary conditions are obviously free (the software  determines them  itself and these conditions are later  chosen at the base of the superwind, except the mass loss fixed to  $\dot{M} \sim 10^{-4} M_\odot/year$.  The conservativity  of the magnetic flux is simply checked.  The steady solution has been found by  trial  and error up to convergence starting  from rescaled preceding results (Pascoli, Lahoche, 2010). The figures 5.1, 5.2 and 5.3  supply respectively the density, the velocity field and the magnetic field in the transition  zone.

\noindent The velocities at the outlet of the nozzle ($r \sim 3\    10^{13} \ cm$)  are moderately high $\sim 35$ $kms^{-1}$. They  are found to be $16$ percent above the liberation velocity ($30 \ km s^{-1}$). We can see that both the gas density and the frozen-in magnetic field intensity  very rapidly decrease on a scale height  $\sim 3.2\ 10^{12}\  cm$. The   area under examination acts similarly to an usual MHD nozzle where the
kinetic energy increases at the expense of the magnetic energy, the flowing
gas going  from a super-alfvenic regime to a sub-alfvenic one. The gradient of density is severe  in this area leading to a strong turbulence and any kind of instabilities. On the other hand the velocities are large and the expansion is adiabatic. The mean temperature decreases  from $\sim 4.5\ 10^4  \ K$ and reaches  $\sim 100-200  \ K$ at the the base of the superwind. The gas emitted at the base  of the superwind  is thus cold as due to the adiabatic expansion.

\noindent  We have  shown in the present study that at least a  steady and laminar solution is existing. However a gentle laminar flow is an indealized situation, a lot of  work remains to be done to understand how exactly  the magnetized disk pierces the  stellar envelope and reaches  the surface. A  time-dependent model  is needed for that but the main difficulty is the treatment of the  numerical instabilities (given the very high  degree of   stratification in  density).

\noindent  There also exists  a lot of physical instabilities which appear when   a magnetic field is present : the magnetic field can for instance possibly produce  in this area  a delicate  small-scale filamentary structure  in the form of twisted ropes (these twisted ropes indeed are   ubiquitous and have been  identified at the surface of the Sun in the corona, but also  in many parts of the solar system, the ionosphere of planets, etc, even though their origin is still controversial. These twisted ropes can form by shearing motion, convergence flow, etc, see for instance,         Priest and Longcope,  2017). There too  a 3D analysis is needed, however we are insured that it is under  this form that the magnetic field  appears at  the base  of the superwind.

\noindent We see very often in the literature that the authors conclude from their observations that  the  (mean)  magnetic field is of the order of $\sim 10 \ Gs$ at the surface (without quotation marks)  of an evolved AGB star. In fact this point of view must be corrected. At the surface of an AGB star the magnetic field is very likely near zero. The intensity deduced from the observations is that  taken at the base  of the superwind (the   top of the "surface" of the evolved  AGB star with quotation marks). Nucci and Busso (2014) use a  mean intensity  $\gtrsim \ 3 \ Gs$  in their calculations. However  such a mean value seems high. In our calculations we find a mean intensity scattered on the evolved AGB star's surface of the order of $1  \ Gs$. In fact this quantity is fictitious  given the field is not scattered but appears concentrate in an  equatorial belt. In anyway the important item is not the intensity of the magnetic field,  but the magnetic flux loss which is imposed by the capacity of the core region to produce such a flux. Furthermore  the star is  hidden from  view of observers by a surrounding  thick disk where the magnetic field is estimated to $1 \ Gs$.

\begin{center}

\vspace{3pt} \noindent
\begin{tabular}{p{10pt} p{15pt} p{15pt} p{48pt} p{50pt}}

{\centering  
$M_c$
} & {\centering 
$M_{env}$
} & {\centering
$\Omega_c$

$(s^{-1})$
} & {\centering     
$\dot{M}$

$(M_\odot\  year^{-1})$
} & {\centering  $\dot{\Phi}$

$(Mx\  year^{-1})$}

\\

\hline \hline
{\centering  $0.5$} & {\centering $0.5$} & {\centering $0.01$} & {\centering  $ 10^{-4}$} & {\centering  $6\  10^{27}$} \\
\hline
\end{tabular}
\vspace{2pt}

\end{center}

\noindent The ejection duration is very grossly  fixed by the total mass of the evolved AGB's envelope, even though the phenomenon would need a very prohibitive CPU time to be fully visualized from the beginning to the end. 

\noindent According to the present model, with calculated mass loss $\dot{M}$ $\sim  10^{-4}$ $M_{\odot}$ $year^{-1}$ and flux loss $\dot{\Phi}$ $\sim  3\  10^{27}\ Mx \ year^{-1}$ (against respectively $\dot{M}_{\odot}\sim 2-3 \ 10^{-14} \ M_{\odot} \ year^{-1}$ and $\dot{\Phi} _{\odot} \sim  10^{22}-10^{24}\ Mx \ year^{-1} $ 
for the Sun), the duration of the phenomenon amounts to $
\sim $ $5$ $10^{3}$ years. Varying the  parameter $\Omega_c$  by a factor $x$, gives $\dot{M} \sim 10^{-4} \times x^2$  $M_\odot \ year^{-1}$ and $\dot{\Phi} \sim$ $3$ $10^{27} \times  x$  $Mx\  year^{-1}$; then this duration is modified  by a factor $x^2$ for the same total mass of the evolved AGB's envelope.   However the value of $\Omega_c$ is linked to the initial mass of the star and thus to the mass of the evolved AGB's envelope itself.

\noindent \textbf{c. The circumstellar envelope}

\noindent    The area under study extends here from $3\ 10^{13}\ cm$ to $3\ 10^{15} \ cm$. The boundary  conditions are specified at the base of the superwind  assumed to be located at $r=3 \ 10^{13} \ cm$ and derived from the results  of the transition area.  At the base  of the superwind (i.e. $3\ 10^{13} \ cm$)  the boundary conditions are $\rho_w=2\  10^{-12} \ g cm^{-3}$ and $H_w=10\  Gs$. The boundary conditions at $r=3\ 10^{15}\ cm$  are left free.  The figures 6.1,  6.2  and 7.1, 7.2  supply the density and magnetic field in the circumstellar envelope.   As shown in figs. 6, in the circumstellar area surrounding the star, the matter is preferentially ejected in a definite plane with formation of a equatorial thick disk surrounding the star (thikness   $\sim 3\ 10^{13}\ cm$). This region is labelled by 1 in fig. 6.2).  The magnetic field remains relatively concentrated in the
equatorial plane with the development of a thick disk. This disk is then continuously extended  by  a horn toward higher latitudes, the latter one
being then interrelated to the bipolar jet (figs. 6.1 and 6.2)  (see the figures published in the literature : M2-9 (Schwarz et al, 1997), He 2-320 and He 3-401 (Sahai, 2002)). Besides, an analysis of the results now reveals that between $r_*$ and  $10\ r_*$  the superwind is not radial but the matter is strongly deviated toward the higher  latitudes.   The density varies as  $r^{-5/2}$ and the magnetic field as  $r^{-3/2}$ in this area.   Beyond $r \sim 50\  r_*$  the motion is quasi-radial $\rho \sim r^{-2}$ and $H\sim r^{-1}$ and the gas distribution   becomes quasi-isotropic  but with a concentration in the polar regions in the form of long  and very directive  jets (these jets  can however evolve toward  a kind of "snake" by instabiliy). We can thus conclude that the inner anisotropies (disk, barrel structure  and jets) are hidden from the observer view by the outer more spherical regions. This fact could explain  why the circumstellar envelopes appear approximately round at an early stage, even though the ejection process is clearly asymmetrical.

\noindent The modelization of the strictly speaking PN's phase   deserves a further examination. The polar holes labelled by 2 in fig. 6.2 must  play  an  important role in  the morphology of PPNe and PNe. When the degenerate core is exposed after the ejection of the evolved AGB's envelope, the hot gas  pushes the dense  gas of the circumstellar  envelope. The effect of bulldozing  is differential, following the direction taken  into account.  The polar cavities with  very low density  offer  lesser resistance than in the equatorial plane direction  (in this plane a thick torus is formed by compression) and the hot gas can pick   up in them creating two  bubbles on  each side of the equatorial thick  torus with appearance of a typical  bipolar PN.

\noindent  Each  cavity  (north or south) is itself  topped  above by a polar jet (labelled by 4 in Fg. 6.2). This jet is produced by a magnetic striction
effect  as early suggested (Pascoli, 1992; Garcia-Segura et
al, 1999). The confinement  is well visible  in NGC 7009 where the jets are very thin and straight (one jet ends up by an umbrella).  Even though the  present model always implies    the formation of these polar jets, there exists two types on PNe. Some PNe possesses very prominent polar blobs  and  likely built-up  from  compression of polar jets (e.g. NGC 650-1,  NGC 4242, NGC 6826, NGC 7009, NGC 3471/2), while  on the contrary  in others PNe  these blobs are lacking at the extremities of the symmetry  axis   (assuming a prolate configuration : e.g.  NGC 6720 and  NGC  7293 (Pascoli, 1990 a,b))\footnote{There is also the very atypical case of the Red Rectangle nebula which is not   barrel-shaped but strangely cone-shaped.  However in this unusual case a binary star as progenitor could be    clearly involved  in the shaping mechanism.}.   It seems that in some objects  the polar structures are reinforced, and by contrast   in others ones  the blobs are eroded by the fast wind and/or the radiation field emitted by the central star  at an evolved stage.

\noindent At a   circumstellar stage  we see just  the outer regions and the envelope appears round as observed (we can also note that the magnetic field appears very  coherent on a  large  scale).  This  is  contrasting with the delicate structures  which is hidden in the inner regions. Following the considerations above  a  circumstellar envelope  must be  a  very complex object indeed.  Even though the ejection process is continuous (one single ejection is assumed), we  mainly distinguish   five structures  labelled  by  the numbers  reported  in  the figure 6.2:

	- (1) A  disk  surrounding the star 
	
	 - (2)  A barrel-shaped  structure with more or less pronounced filaments 
	 
	- (3) Polar  holes  inside the barrel-shaped structure
	
	- (4) A polar jet  above each  hole  piercing  the barrel-shaped substructure
	
	- (5) A quasispherical envelope in the outer region   which masks the inner structures excepting  the extremities  of the  polar jets.
	
\noindent We can point out that in the present scenario,  all  these structures  derive  by inflation  of  a   magnetized disk emitted by the central star and not from a bipolar  outflow. The bipolarity is shaped after the ejection by long-range Lorentz forces. On the other hand the fast wind emitted  when the degenerate core is exposed is  very likely spherical.
	
\noindent A magnetic field line is  continous and without boundary (a piece of magnetic filament with two extremities is obviously unphysical). It must be a closed  path even if this path is not a single  loop but can be very interlaced. What is the fate of a magnetic filed line in an evolved AGB star-circumstellar envelope system  ? We show in fig. 8   the  full  path of  such a line (only one hemisphere is shown). It goes from the core of the evolved AGB star  to the nebula  by the equator and returns to the core  by the polar axis. This line can draw in the body of the nebula a clelia. 	Let us notice that the magnetic  field entering by the pole in the star is radial ($H_r<0$). By conservation of flux, this signifies that the field in the equatorial  plane is not purely toroidal but possesses a small pitch angle  with $\frac{H_r}{H\varphi}\sim 0.1$ and  $H_r >0$. A piece of  twisted rope  is also shown, undoubtedly built-up in the transition area (sole a small piece is shown but in fact by continuity of the carried-current  it is all the magnetic field line which must exhibit such a  twisted structure).

\noindent If the gas is not homogeneous but filamentary, a very likely situation when a magnetic field is present, a bundle of magnetic lines could be more prominent than others and could   directly be visualized.   Some nebulae seems to  exhibit such substructures accompanied of filamentation effects (Huggins and Manley, 2005). A comparison with a composite image  of NGC 6543 (the famous Cat's Eye nebula)  however suggests that the   polar jet might  not  stay  straight as represented here, but rather  could undergo  a kink  deformation. Thus one of the two symmetric jets in the Cat's Eye nebula is broken at right angle (if this is due to the presence of a magnetic field, we have an     instability in the  azimuthal mode $m=1$). We can observe the same kink effect on a very prominent filament in M2-9.   By the way it seems  difficult to explain  such a  net  symmetry  breaking  by a contiuous  precession of a binary nucleus. We observe a second type  of instability in  the opposite jet which is  divided  into two secondary separate jets (instability in the filamentation  modes $m\geq 2$).  We can also note the strinking resemblance between NGC 6543 (the Cat's eye) and NGC 7009 (the Saturn nebula) : multiple concentric thin shells, small scale filaments and  bipolar  jets.  In the latter one we can see one  of the jets with a thread-like structure ended with a stagnation point  at its tip (to be compared with  the figure 4 of Bellan, 2018). There also exists   a  very conspicuous and intriguing  braid  composed of thin twisted ropes well  visible in a lobe of M2-  9, a phenomenon  which is very difficult to explain without the presence of a magnetic field but easy with it.  Some types of these plasma  instabilities are described in Ciardi et al (2009), Bellan (2018) and papers quoted therein, even though an extrapolation of laboratory experiments to  astrophysical objects  must be taken with cautious,  and  the very  large scale difference (especially concerning the   characteristic time scale for the magnetic diffusivity) must be keep in mind.  

\noindent For the  very complex and archetypal nebula M2-9 which seems to bring together  all the difficulties, other interpretations   has also been proposed with both a binary star progenitor  and periodic  mass ejections (Balick,  Wilson, Hajian, 2001). There is a remarkable series of outer, evenly spaced, spherical shells of gas. The duration  between each ejection (in the framework of a multiple shell ejection) is rather surprisingly  stable and of the order of $1000\ years$. However this period is very different from that linked to  thermal pulses $\gtrsim 10^4 \ years$ (Lau,  Gil-Pons, Doherty,  Lattanzio, 2012) or to pulsations of evolved AGB stars $\sim \ 1 \ year$ (McDonald,  De Beck,  Zijlstra and Lagadec, 2018). The apparent absence of turbulence in these concentric envelopes is also troublesome.    A simple suggestion is  that this turbulence could  be cut off by a large-scale magnetic field in the short wavelength range and the energy generated by  a  shock   would be then  stored in  wavelengths of the order of the distance between the shells. However this idea is  very difficult to treat from  a theoretical point of view. Another idea will be to see this series of concentric envelopes as a soliton train, well known in many contexts such as undular bores and whelps in hydrodynamics and meteorology,  where the  magnetic striction would play the role of the usual gravity, completely negligible here.  More generally,  in  light of the present model, we  hypothesize that this  multiple  shell structure  could   possibly have an   in situ  origin in the spherical envelope (the descendant of the round circumstellar envelope surrounding the barrel-shaped  substructure).  Magnetic instabilities can directly act  within the  spherical envelope, all the  phenomena taking place  in the framework of a continuous and unique flow of gas  without appeal to an abrupt change in the mode of ejection.  On the other hand  the geometry of this strangely sliced spherical envelope  exhibits  a very impressive  contrast with the inner region which is clearly  strongly axisymmetric. A contrario  in the scenario of a multiple ejection  process what could be the  cause of    this  very sudden  regime change from a near perfect spherical mode to an  axisymmetric one during the superwind phase without any change of the gas chemical composition ?

\noindent Let us eventually notice that the Cat's Eye  is also surrounded by an  enormous spherical (but very faint)  envelope (Corradi, 2004).  This outer envelope is composed  of a complex network of interlaced filaments  whose the morphology presents  strong similarities with  the Crab Nebula  where a magnetic field have been clearly identified ($\sim 10^{-4}-10^{-3} \ Gs)$), and this even though the  expansion velocities  in these two  distinct  objects (a planetary nebula for the Cat'Eye with no   synchrotron emission and a remnant of supernova for the Crab nebula with   synchrotron emission)  differs  by a factor $100$.

\hspace{-30pt}\includegraphics[height=200pt,width=250pt]{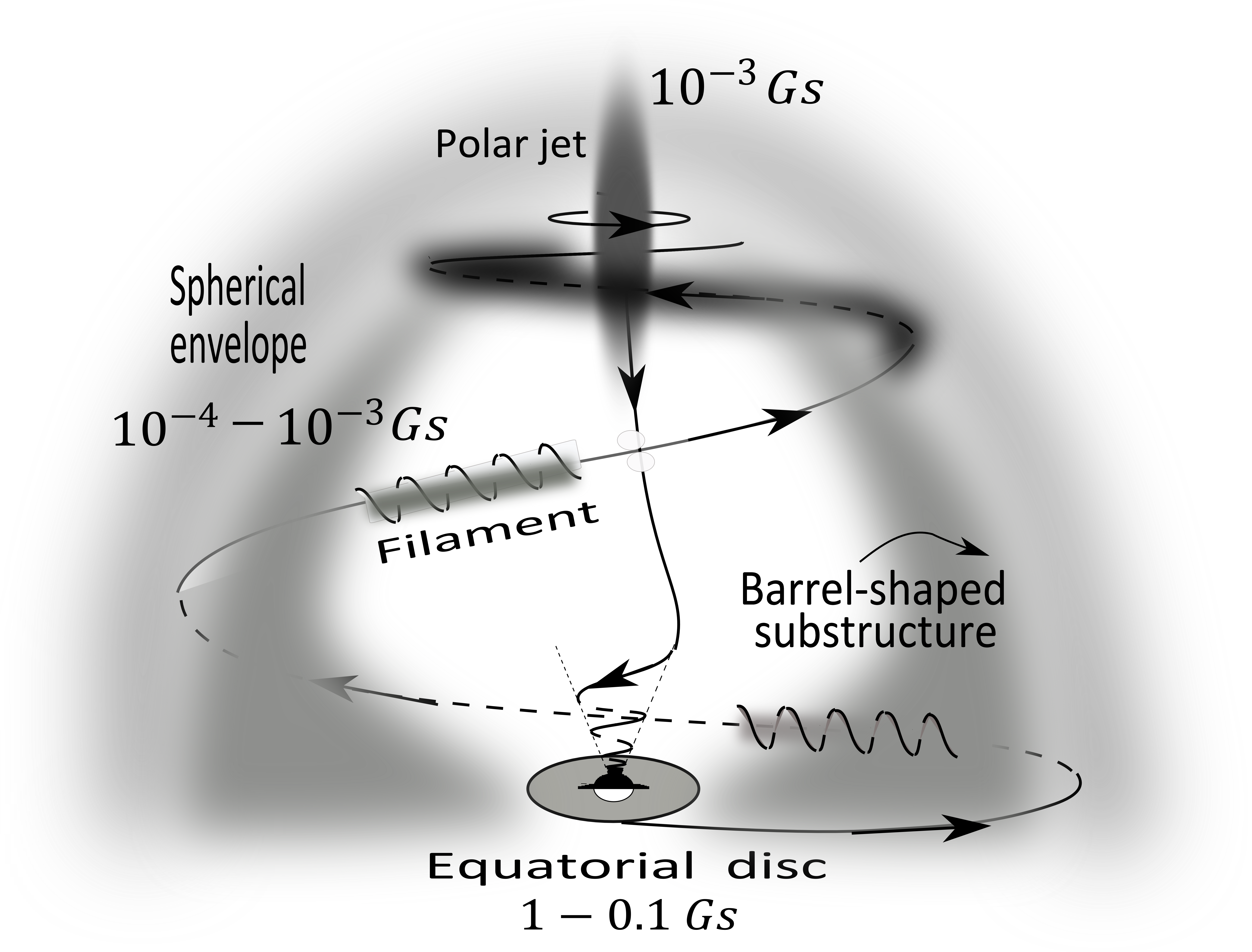}

\noindent Figure  8  : Synoptic view  of a magnetic bundle in a circumstellar envelope. The barrel-shaped structure is hidden by a quasi-spherical envelope (sketch from  fig. 6.2). The core is also reported (not to scale). Circumstellar envelopes evolve in PNe which keep still track   of their origin, see by comparison NGC 3242, NGC 6826, NGC 7009 and NGC 6843 from the Hubble Telescope. In NGC 7009 we can  see the confined jets (the so-called streams in Sabbadin et al, 2004) and their clear links with  the barrel-shaped  structure.

\noindent Eventually  as we have already said above the big problem is  that a simulation model is very-time consuming and costly. A semi-analytical model would be welcome (a bit like  that of Nucci and Busso (2014), however these authors did not consider  either the dynamo mechanism, or the ejection process by the evolved AGB star. A partitioning of the study in different areas   would be needed in this case.

\section{Conclusion}

We  have succeeded in producing a self-consistent time-dependent MHD model of circumstellar envelope  ejection at the tip of the  AGB phase. The present paper supplies  new results     compared to the preceding time-independent  models of Pascoli and Lahoche (2010). We start with  a spherical AGB star composed of  a very huge convective envelope surrounding  a small and dense degenerate core. After incorporating  an angular velocity distribution, the  symmetry is lowered in the core region, going from spherical to cylindrical.   A meridional circulation then naturally appears. This circulation in  turn produces a reshuffling of the angular velocity wich becomes polar. We predict the existence of a huge polar vortex above the degenerate core of an evolved AGB star. Above this  degenerate core, fast angular velocities are then located at higher  latitudes where a dynamo is operating.  Magnetic fields are then amplified and stored away from the dynamo region toward the equatorial plane where a thin magnetized disk develops. A magnetic field  of the order of $10^6\ Gs$ is predicted just above the core. The disk rises later radially throughout the evolved AGB's envelope up to the "surface" by buoyancy. The magnetic field value at the "surface" is estimated at $10 \ Gs$ (the measurement  is made in the zone where the wind is emitted, this giving  a mean magnetic field (a fictitious  value)  at the surface of the AGB of $\sim\ 1\ Gs$. This ambiguity present in a lot of  papers  does  that the concept  of intensity should give way to that   of    magnetic flux which is much more relevant).  In this framework the ejection of gas by the AGB star at its "surface" is MHD-driven. The mean magnetic flux  loss and the expansion velocity  are predicted in good agreement with observational inferences ($\dot{\Phi} \sim\  3\ 10^{27}$ $Mx\ year^{-1}$), $v_e \sim 20$ $kms^{-1}$. The mean mass loss is more difficult to determine, but a mass loss $\dot{M} \sim 10^{-4} M_\odot/year$ is easily obtained, comparable to the observational estimates. However much higher mass loss could also be produced (as in the case of M2-9).  The transition from a grossly round circumstellar envelope toward the characteristic bona fide bipolar morphology of  PNe has also been questioned. This noticeable morphology could result at a post-AGB stage, when the degenerate core is exposed. The shaping of the bipolarity is achieved starting from the hidden anisotropic distribution of matter buried  in the inner region of the circumstellar envelope.  This question is left as a further work. Eventually to understand the surimposed filamentary structure,   a  3D model taken into account of an azimuthal dependence for the physical quantities is required, but will need a huge amount of  CPU ressources. More generally many  interrogations  relative to  the  existence of  magnetic fields in the circumstellar envelopes of evolved AGB stars and in PNe remain to  be answered.

\begin{center}
\includegraphics[height=250pt,width=278pt]{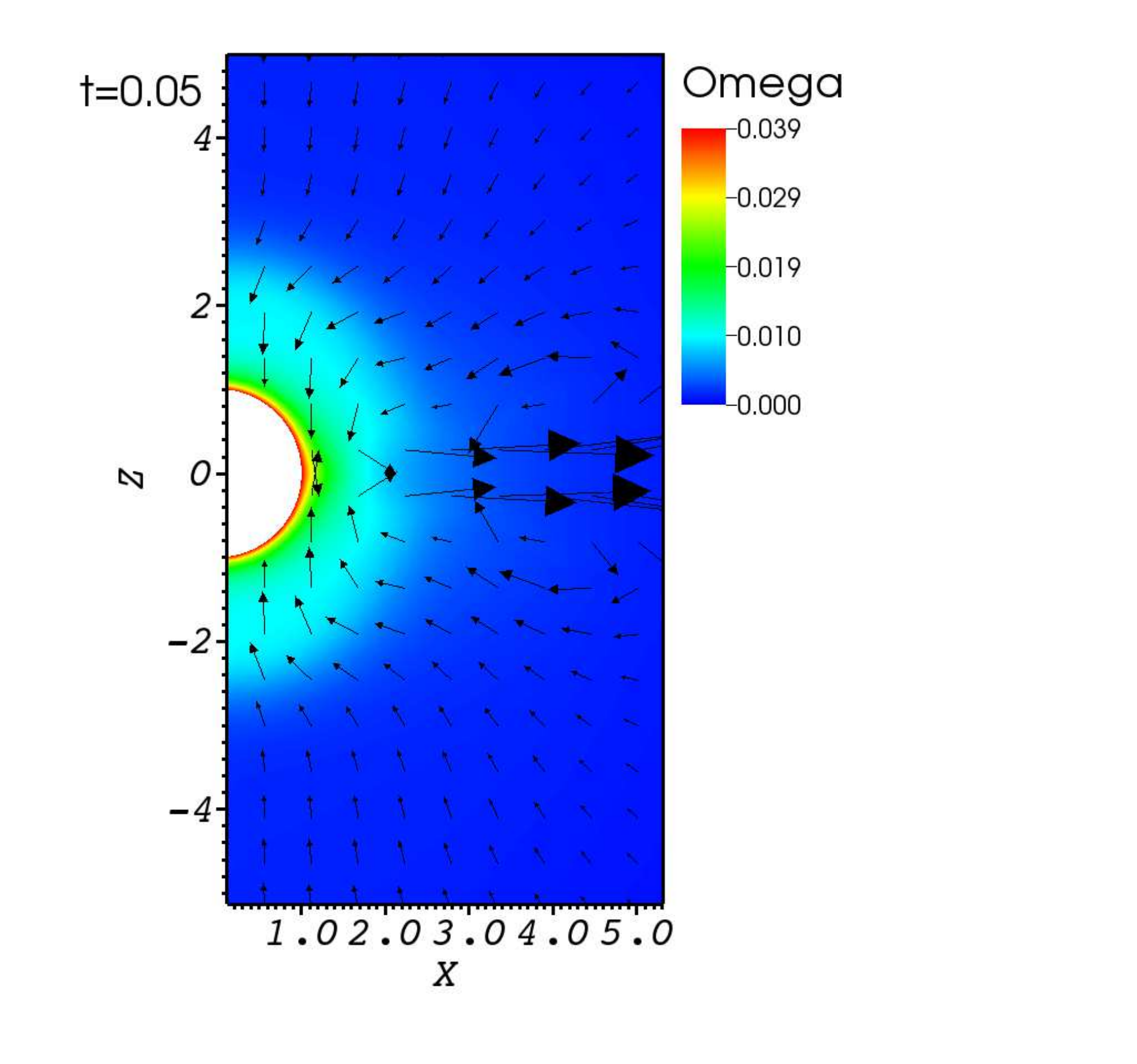}

Figure 1.1
\end{center}

\begin{center}
\includegraphics[height=250pt,width=278pt]{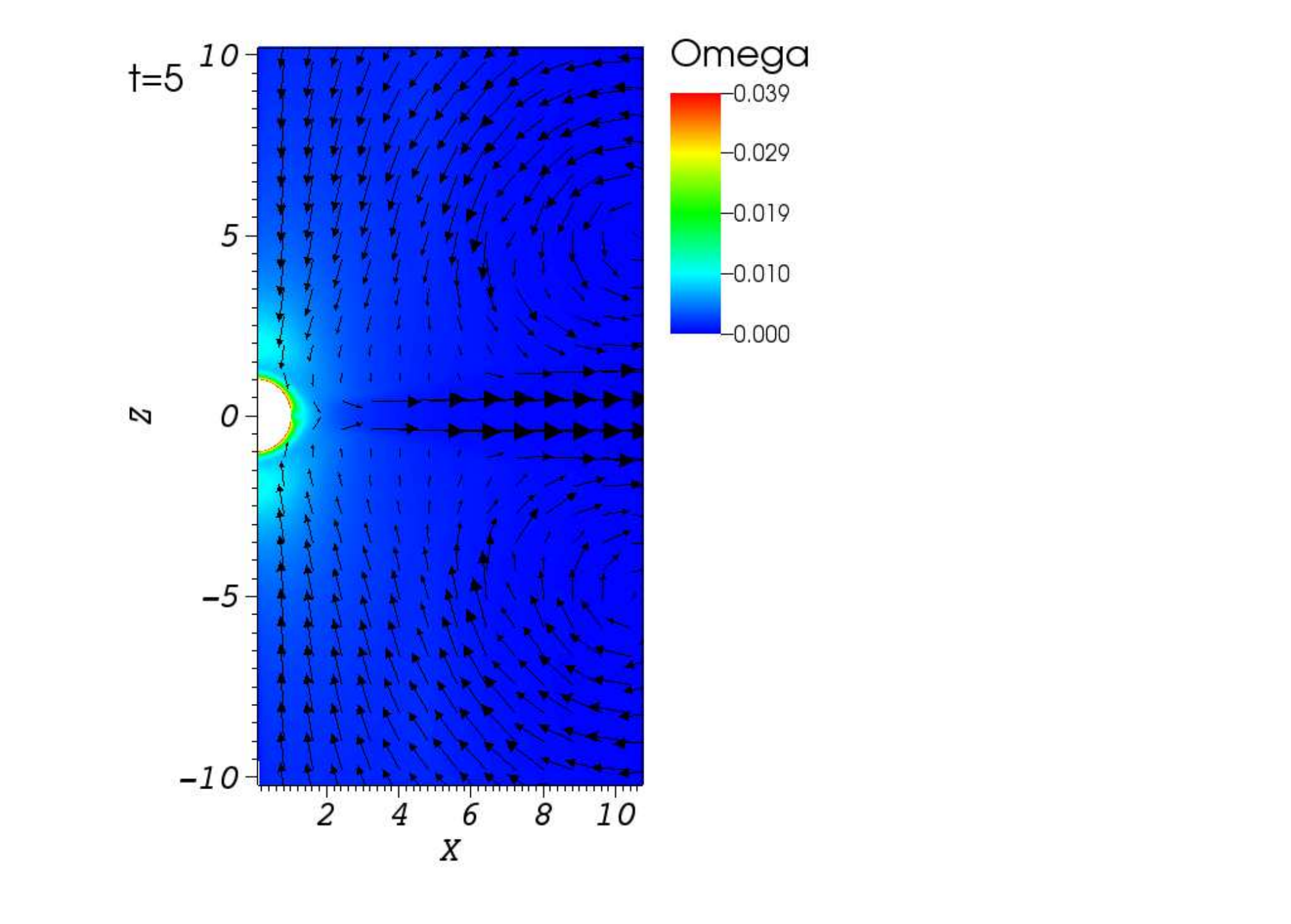}

Figure 1.2
\end{center}

\begin{center}
\includegraphics[height=250pt,width=278pt]{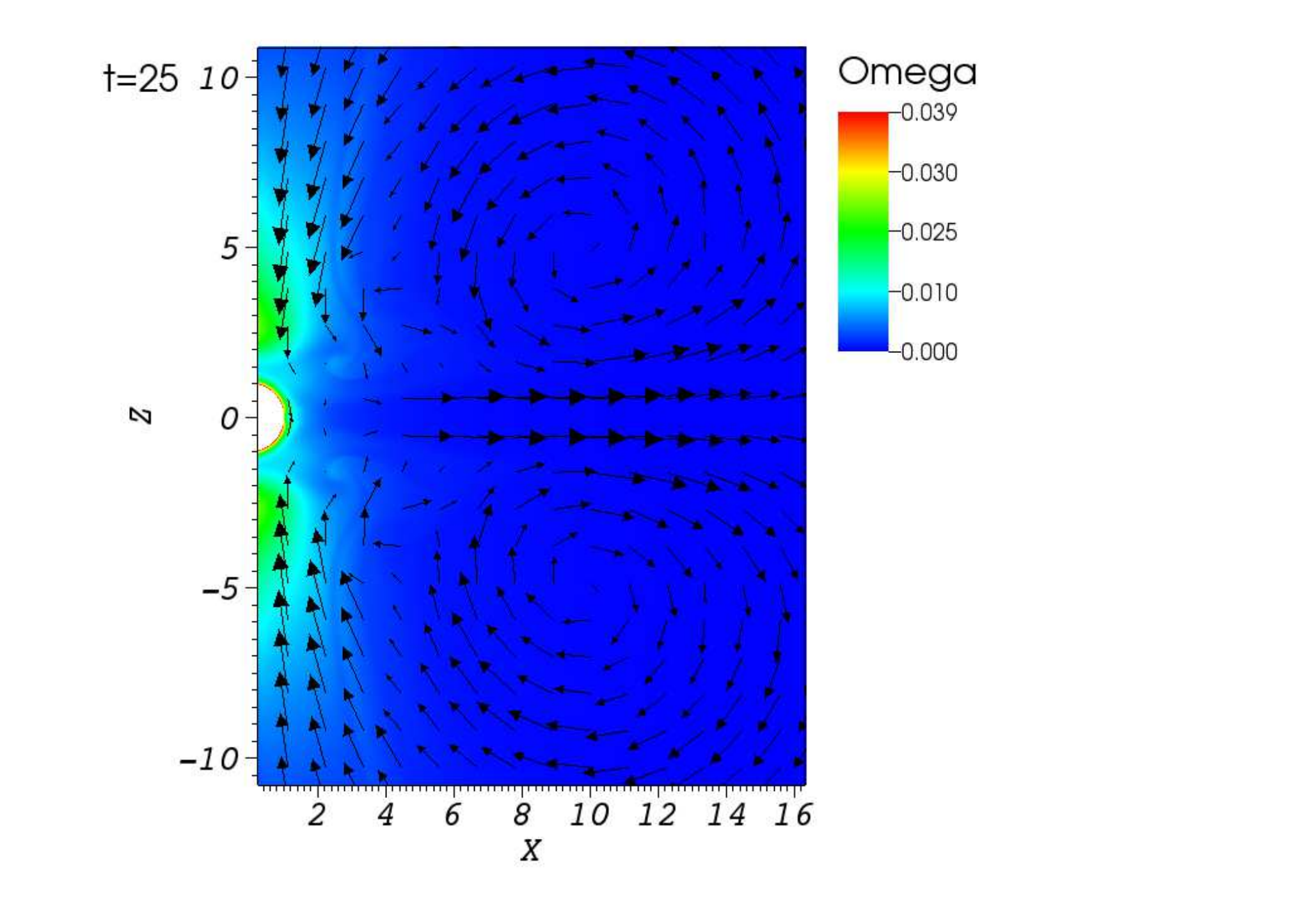}

Figure 1.3
\end{center}

\noindent Figure 1 Core region : Angular velocity in normalized units, unit $=3 \ 10^8 \ km s^{-1}$ (log scale). The meridional velocity field is surimposed (characteristic velocities  $\sim 5\ km s^{-1}$). The time unit is  $5  \ h$. 

\begin{center}
\includegraphics[height=210pt,width=228pt]{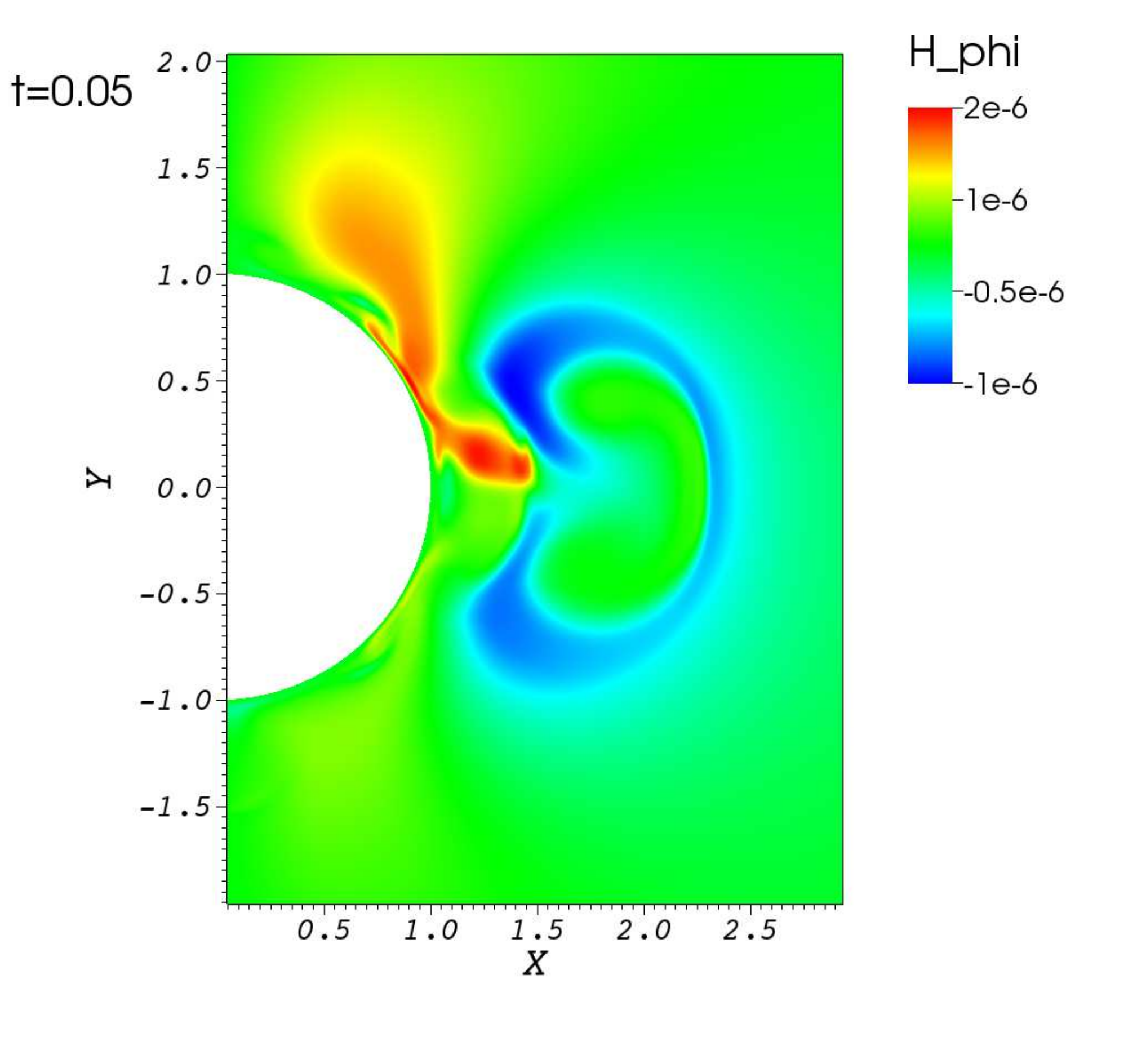} 
Figure 2.1
\end{center}

\begin{center}
\includegraphics[height=210pt,width=228pt]{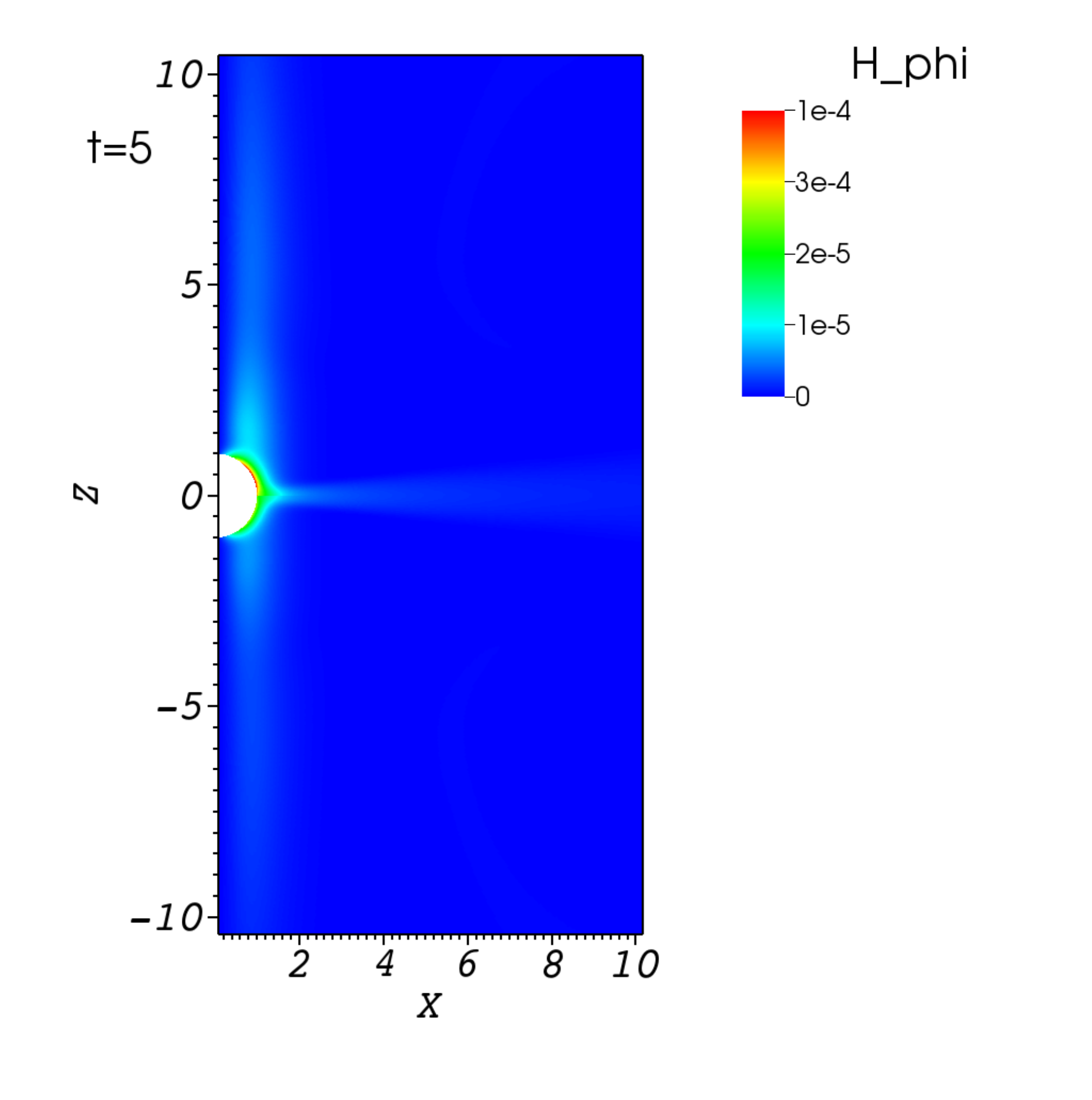} 
Figure 2.2
\end{center}

\begin{center}
\includegraphics[height=210pt,width=228pt]{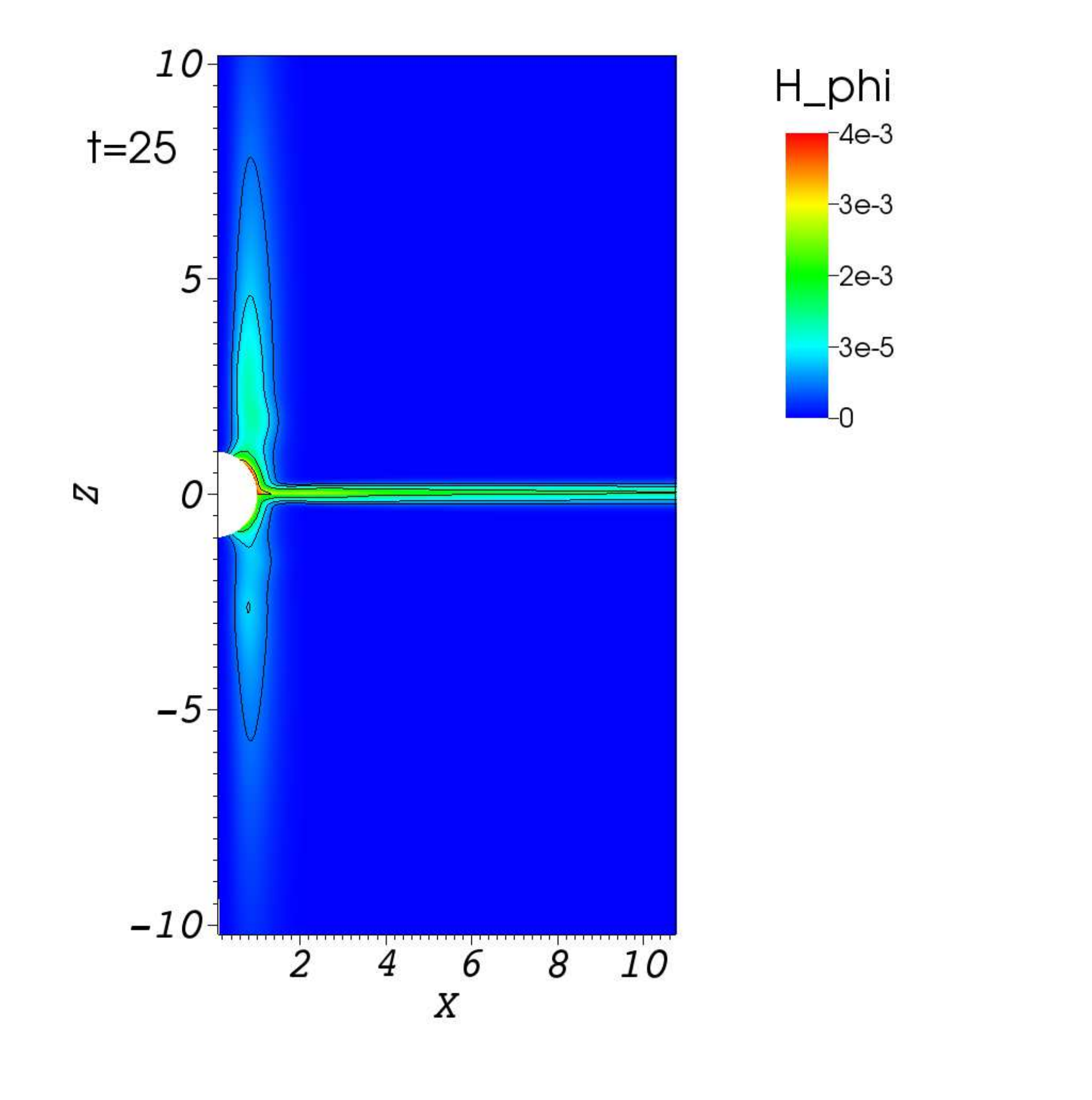}
Figure 2.3
\end{center}

\noindent Figure 2 Core region :  Azimuthal magnetic field in normalized units, unit $=7\ 10^8 \ Gs$  (fig. 2.1 linear scale, figs 2.2, 2.3 log scale). The time unit is  $5  \ h $.

\begin{center}
\includegraphics[height=210pt,width=228pt]{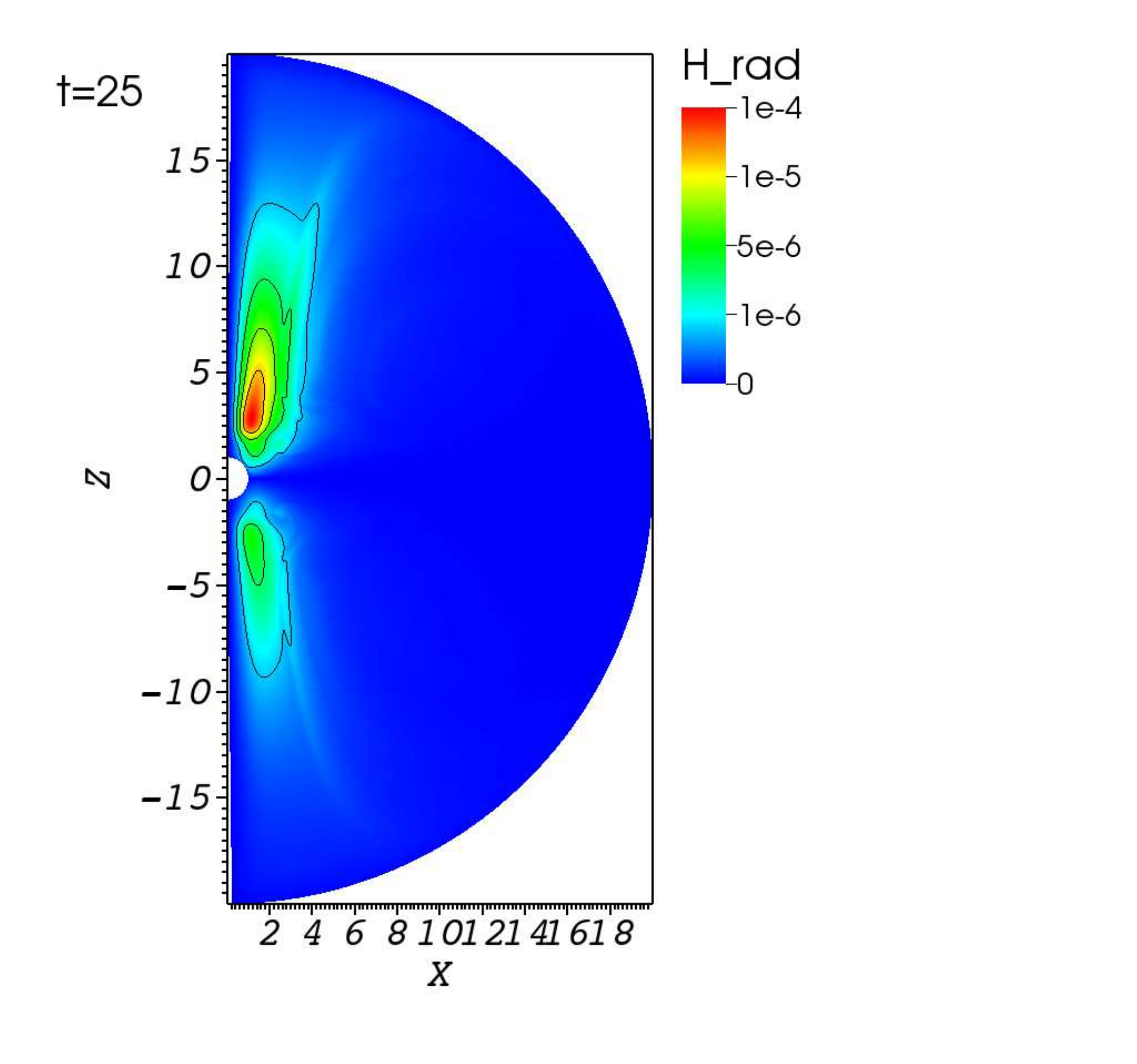} 
\end{center}

\vspace{-20pt}
\noindent Figure 3 Core region :  radial magnetic field in normalized units, unit $=7\ 10^8 \ Gs$ (log scale) 

\begin{center}
\includegraphics[height=210pt,width=228pt]{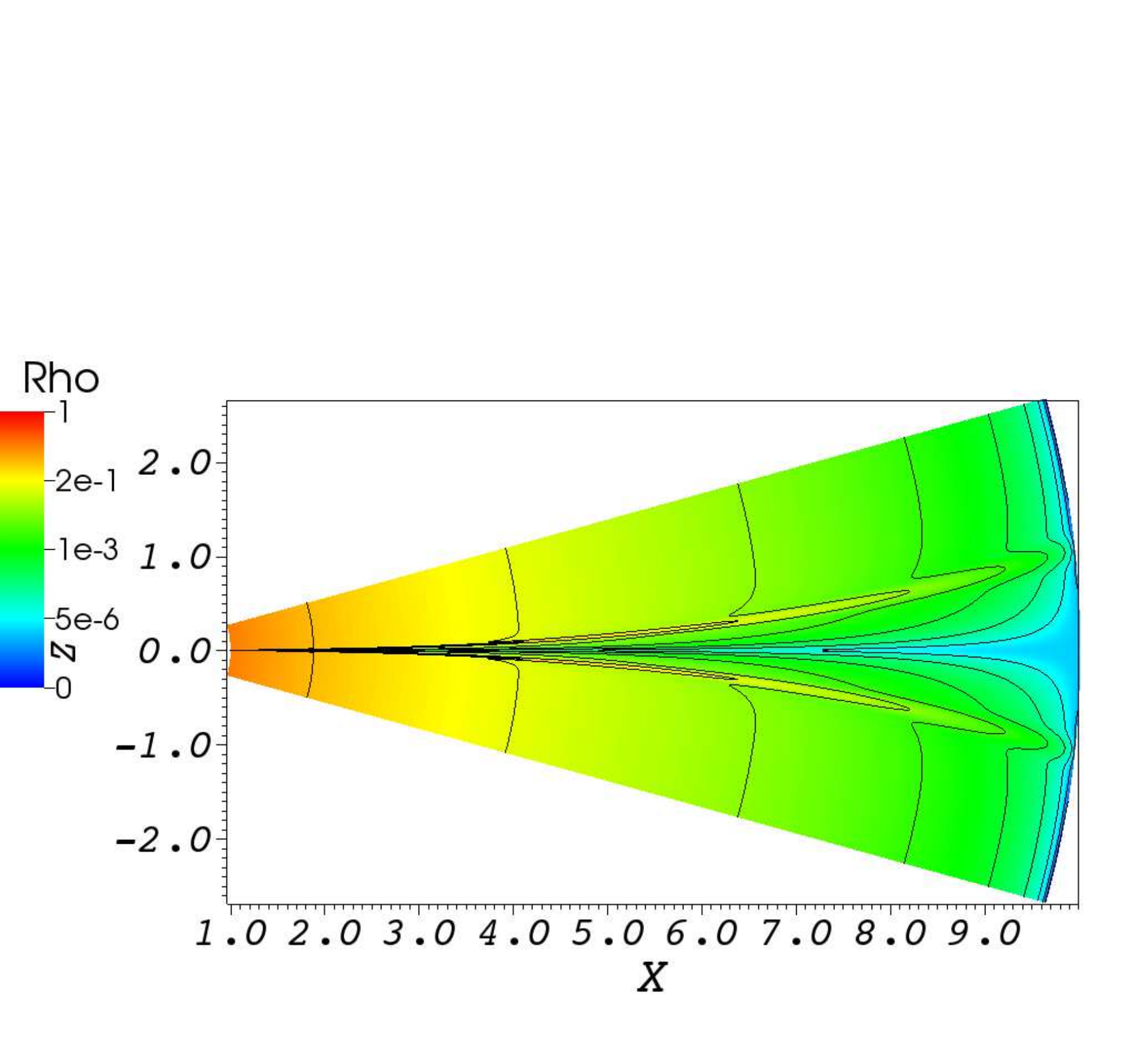} 
\end{center}

\noindent Figure 5.1 Star-superwind transition area   : Surface and contour plots of the density $\rho$ in normalized units, unit $= 5.3  \ 10^{-7} \ gcm^{-3}$ (log scale).  The unit of distance is  $3\ 10^{12} \ cm $. 

\begin{center}
\includegraphics[height=210pt,width=228pt]{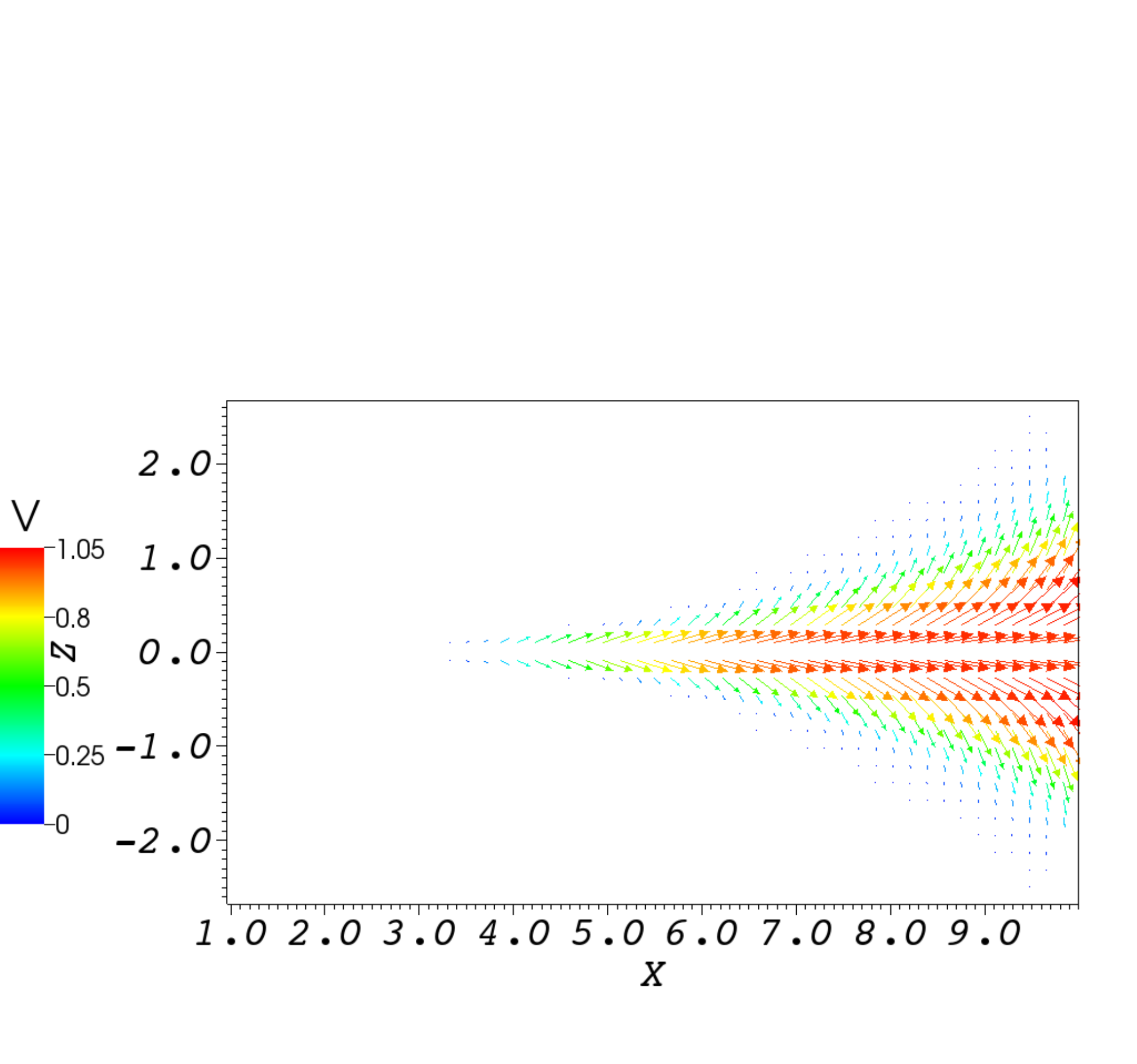} 
\end{center}

\vspace{-25pt}
\noindent Figure 5.2 Star-superwind transition area  : Velocity field in normalized units (unit $=30\  km \ s^{-1}$).  

\begin{center}
\includegraphics[height=210pt,width=228pt]{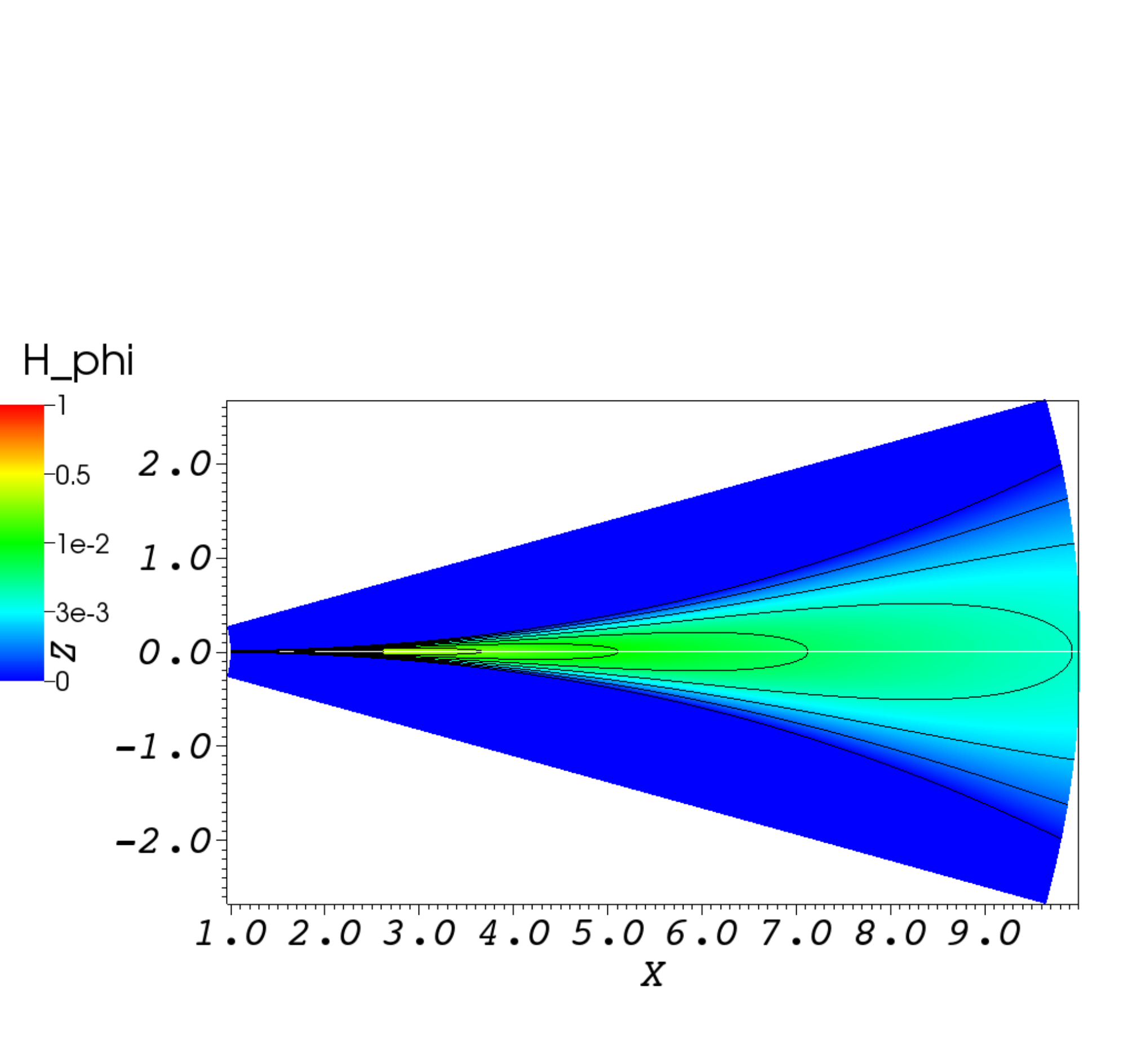} 
\end{center}

\noindent Figure 5.3  Star-superwind transition area : Surface and contour plots of the magnetic field  $H_\varphi$ in normalized units, unit $=2.3\ 10^3 \ Gs$ (log scale). 

\newpage 

\begin{center}
\includegraphics[height=210pt,width=228pt]{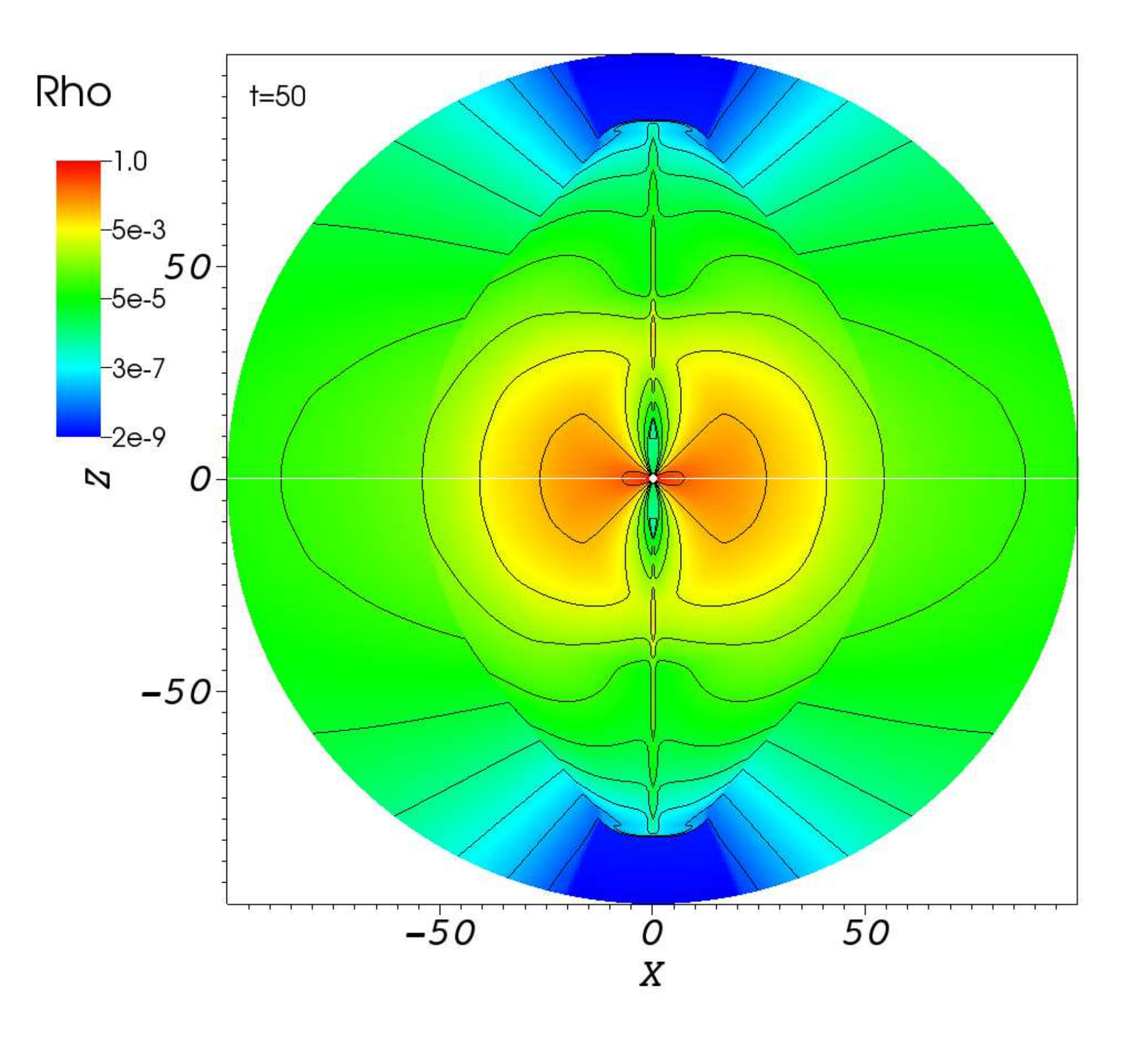} 
Figure 6.1
\end{center}

\begin{center}
\includegraphics[height=210pt,width=228pt]{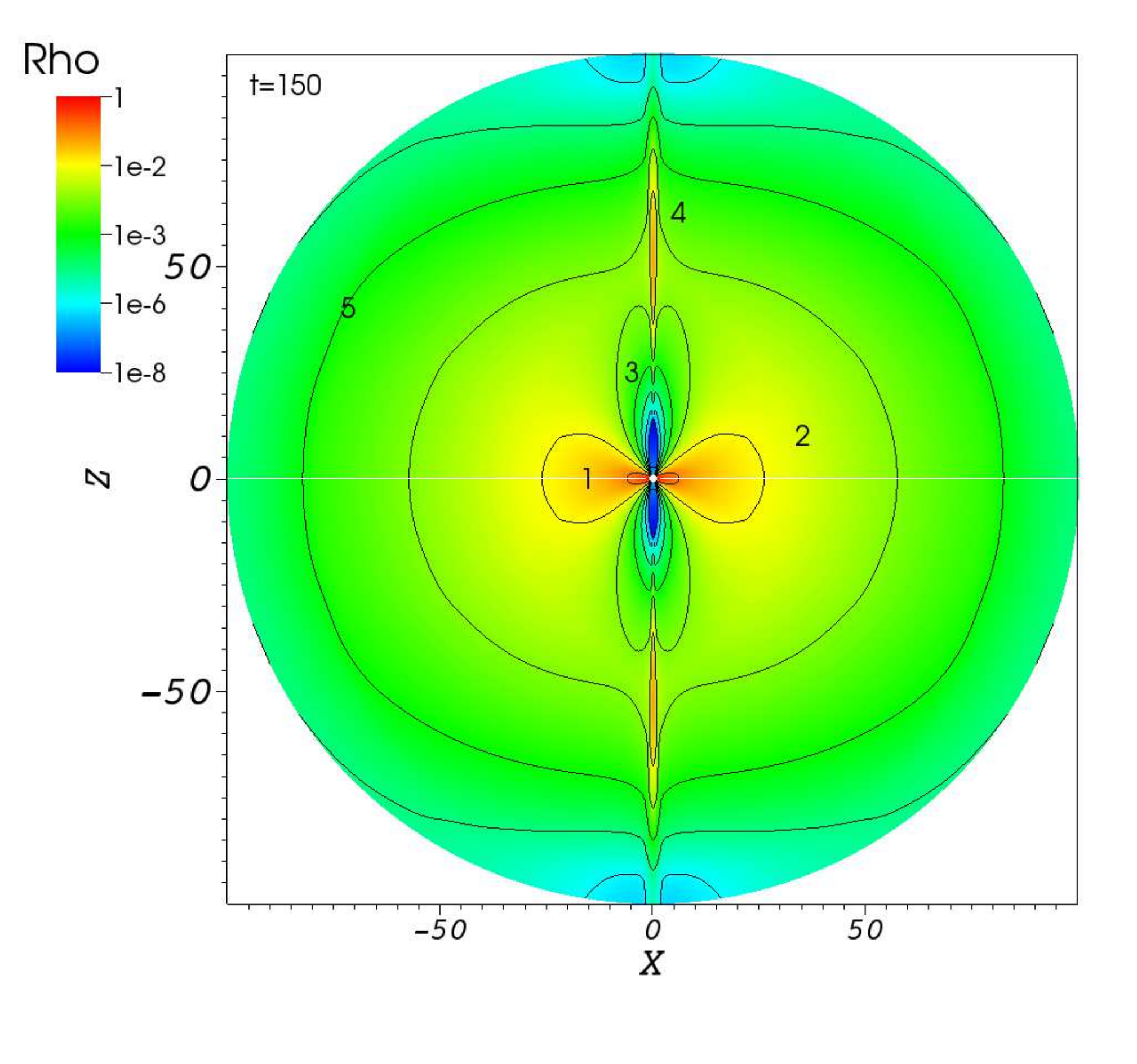} 
Figure 6.2
\end{center}

\noindent Figure 6  Circumstellar envelope: Surface and contour plots of the density $\rho$ in normalized units, unit $=2\ 10^{-13} \ gcm^{-3}$ taken in the disk  (log scale). The unit of distance is the radius of the star $r_*=3\  10^{13} \ cm $. The time unit is  $100 \ d $.

\begin{center}
\includegraphics[height=210pt,width=228pt]{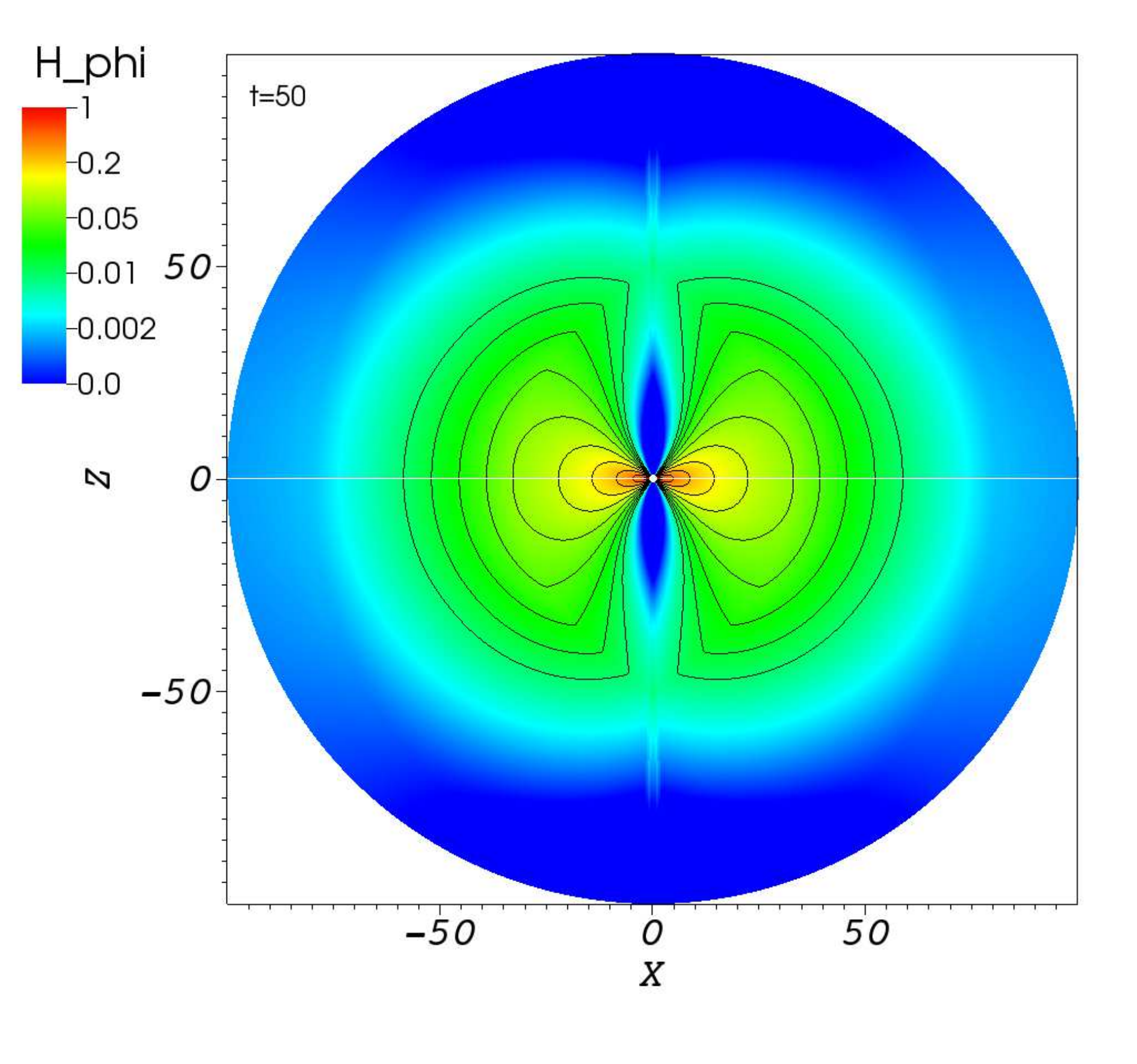}

Figure 7.1
\end{center}

\begin{center}
\includegraphics[height=210pt,width=228pt]{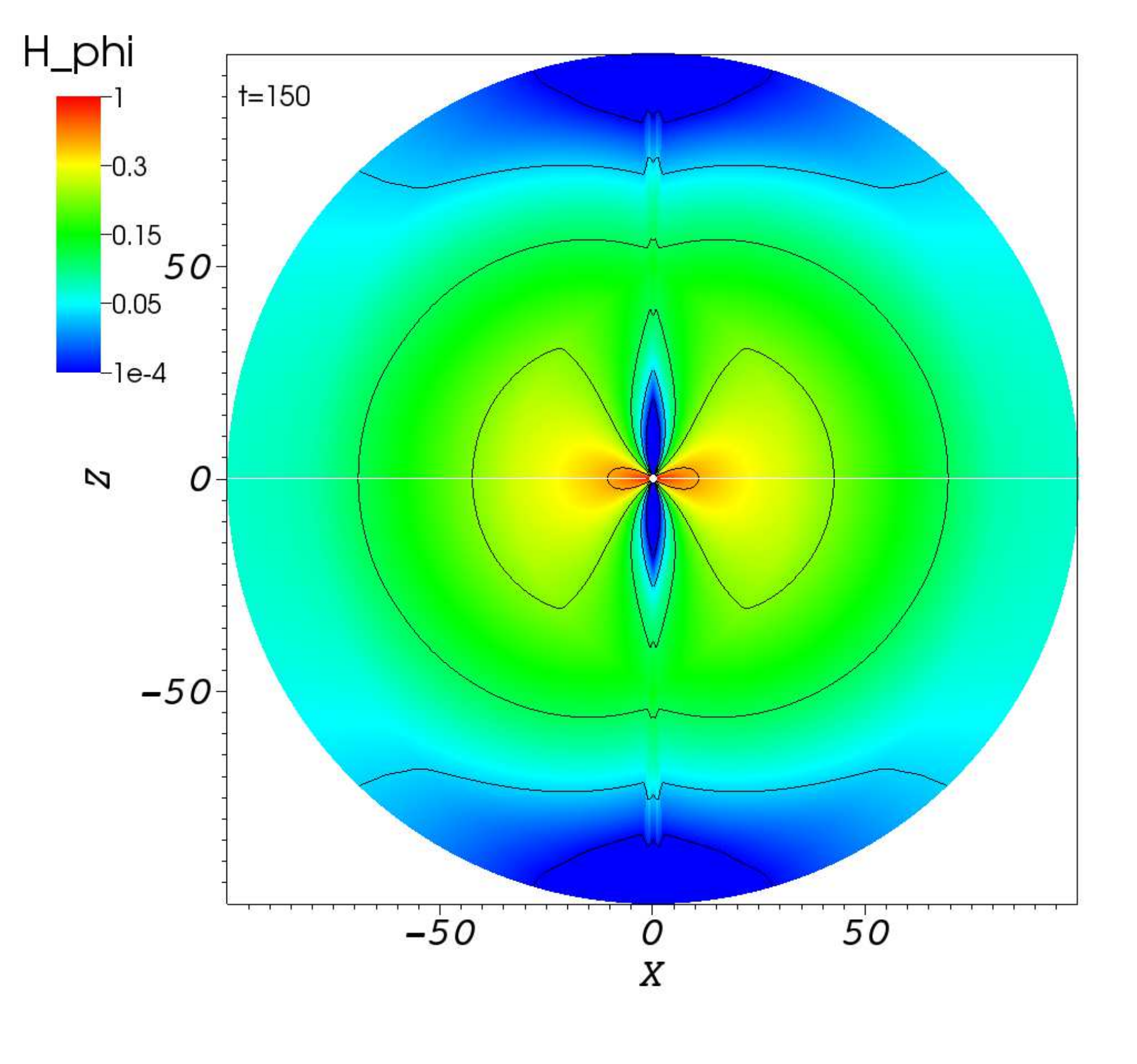}
Figure 7.2
 \end{center}

\noindent Figure 7  Circumstellar envelope : Surface and contour plots of the magnetic field $H_\varphi$ in normalized units, unit $=1 \ Gs$ taken in the disk (log scale). The unit of distance is the radius of the star $r_*=10^{13} \ cm $. The time unit is  $100 \ d $. 

\begin{raggedright} 
\textbf{Acknowledgments :}
\end{raggedright} 

\noindent The numerical results presented here were obtained using the ressources of the MeCS platform of the Universit\'{e}  de Picardie Jules Verne.

\noindent  The author would like to thank an anonymous referee for his   insightful comments.

\begin{center}
\textbf{References}
\end{center}

   Balick, B., Wilson, J.,  Hajian A.R., Astron. J., 2001,  121, 354

 Beck, R., Brandenburg, A., Moss, D.,  Shukurov, 1996,
Annu. Rev. Astron. Astrophys, 34, 155

 Bellan, P.M., 2018, Plasma Physics,   25, 055601

 Blackman, E.G., Franck, A., Markiel, J.A., Thomas, T.H.,
Van Horn, H.M.,  2001, Nature, 409, 485

 Blackman, E.G. 2009, Cosmic Magnetic Fields: From
Planets, to Stars and Galaxies, Proceedings of the International
Astronomical \ Union, IAU Symposium, Volume 259, p.35

  Brandenburg, A., Kleeorin, N.I.,  Rogachevskii,I., 2016, New J.  Physics  18, 125011

 Busso, M., Gallino, R., Wasserburg, G., 1999, ARAA 37, 329

 Cabanes, S.,  Schaeffer, N., Nataf, H.C., 2014, phys. Rev. Letters, 113, 184501

 Charbonneau, P.,  2010, Living Rev. Solar Phys., 7, 3

 Chevalier, R.A., Luo, D.,  1994, ApJ, 421, 225

 Ciardi, A., Lebedev, S.V., Frank, A., Suzuki-Vidal, F.,
Hall, G., Bland, S.N., Harvey-Thompson, A., Blackman, E.G., Camenzind, M.
2009, ApJ Letters, 691, L147

 Chatterjee, P.,Nandy, D, Choudhuri, A.R. 2004, A\&A,
427, 1019

 Corradi,  R., 2004, https://www.spacetelescope.org  /images/heic0414b/

 Dikpati, M., Choudhuri, A.R., Sch\"{u}ssler, M. 1995,
A\&A, 303, L29

 Dikpati, M., Charbonneau, P., 1999, ApJ, 518, 508

 Dobler, W., Stix, M., Brandenburg, A., 2006, ApJ, 638, 336

  Duthu A.,  Herpin, F.,  Wiesemeyer, H.,  Baudry, A.,  L\`{e}bre, A., Paubert, G., 2017,   A\&A,  604, A12    

 Ferrario, L., de Martino, D.,  Gaensicke, B., 2015,  Space Science Reviews. 191, 111

 Fossat, E.,  Boumier, P.,  Corbard, T., Provost, J., Salabert, D.,  Schmider, F.X.,  Gabriel, A.H.,  Grec, G., Renaud, C.,  Robillot, J.M., Roca-Cortés, T., Turck-Chièze, T.,  Ulrich, R.K.,  Lazrek, M., 2017, A\&A, 604, A40 

 Garc\'{\i}a-Segura, G.,, L\'{o}pez, J.A., Franco, J. 2005,
ApJ, 618, 919

 Garc\'{\i}a-Segura, G.,, Langer, N., R\'{o}yczka,
M.,Franco, J. 1999, ApJ, 517, 767

 Gilman, P.A., Rempel, M.,  2005, ApJ, 630, 615

 Guerrero, G.A., Mu\~{n}oz, J.D., de Gouveia Dal Pino,
E.M. 2005, Magnetic Fields in the Universe, From Laboratory and Stars to
Primordial Structures, AIP Conference Proceedings, 784, 574

 Gurzadian, G.A.,1969, Planetary Nebulae, Gordon and
Breach, New York

  Habing, H.J., 1996, A\&ARv, 7,  97

 Herpin, F.,  Baudry, A., Thum, C., Morris,D., Wiesemeyer, H.
2006, A\&A, 450, 667

  Herwig, F.,  2005,  ARA\&A, 43, 435

 H\"{o}fner S., Olofsson H., 2018, A\&ARv, 26, 1  

 Huggins, P.J., Manley, S.P.,  2005, PASP, 117, 665

 Jordan, S., Werner, K., O'Toole, S.J., 2005, A\&A, 432, 273

 Jordan, S., Bagnulo, S.,  Werner, K., O'Toole, S.J.,  2012, A\&A, 542,  A64

  Karakas, A.I.,  Lattanzio, J., 2014, PASA, 31, e030

  Kemball, A. J., Diamond, P. J., Gonidakis, I., et al. 2009, ApJ, 698, 1721

 Kitchatinov, L.L., 2016, Geomagnetism \& Aeronomy, 56, 945

  Kerschbaum, F., Ladjal,  D., Ottensamer, R., Groenewegen, T., Mecina,
M., Blommaert, J.A.D.L., Baumann,  B., Decin, L.,  Vandenbussche,  B.,   
Waelkens, C.,  Posch, T.,  Huygen, E.,   De Meester, W., Regibo, S., Royer, P., Exter, K. 
and  Jean, C.,  2010, A\&A,  518, L140

 Krause, F., R\"{a}dler, K.-H. 1980, Mean-field
Magnetohydrodynamics and Dynamo Theory, Pergamon Press, Oxford

 K\"{u}ker, M., R\"{u}diger, G., Schultz, M. 2001, A\&A 374,
301

 Lau, H.H.B.,  Gil-Pons, P., Doherty, C.,  Lattanzio, J., 2012, A\&A, 542, A1

  L\`{e}bre, A., Auri\`{e}re, M., Fabas, N.,  Gillet, D.,  Herpin, F.,  Konstantinova-Antova R. and  Petit., P.,  2014, A\&A, 561, A85

    Liang, Z-C.,  Gizon, L.,   Birch, A.C.,   Duvall Jr, T.L.,  Rajaguru, S.P., 2018, A\&A, 619A,  99

 McDonald, I., De Beck, E., Zijlstra A., Lagadec, E.,  MNRAS,   2018, 481, 498

 Matt, S., Balick, B., Winglee, R.,Goodson, A., 2000, ApJ., 545,965

 Mignone, A., Bodo, G., Massaglia, S., Matsakos, T. Tesileanu, O., Zanni, C., Ferrari, A.,  2007, ApJS, 170, 228

 Mignone, A., Zanni, C., Vaidya, B., Matsakos, T., Muscianisi, G., Tzeferacos, P., Tsileanu, O., PLUTO User's Guide, v.4.2, 2015

 Neri, R., Kahane, C., Lucas, R., Bujarrabal, V.,  Loup, C., 1998, A\&AS, 130, 1

 Nordhaus, J.T., Blackman, E.G., Franck, A., 2007, MNRAS,
376, 599

 Nordhaus, J., Busso, M., Wasserburg, G. J., Blackman, E. G., Palmerini, S., 2008, ApJ Letters, 684, L29

 Nucci, A.M., Busso, M.,  2014, ApJ,  787, 141

 Paternò, L., Sofia, S. Di Mauro, M.P., Astron. Astrophys. 314, 940

 Parker, E.N., 1984, ApJ, 281, 839

 Pascoli, G., 1987, A\&A, 180, 191

 Pascoli, G., 1990a, A\&A, Suppl Series, 83, 27

 Pascoli, G., 1990b, A\&A, 232, 184      

 Pascoli, G., 1992, PASP, 104, 305

 Pascoli, G., 1997, ApJ, 489, 94

 Pascoli, G., Lahoche, L., 2008, PASP, 

 Pascoli, G., Lahoche, L., 2010, PASP, 120, 1267

  Pipin, V.V.,  Kosovichev, A.G. 2011, 738, 104

  Priest, E. R.,  Longcope, D. W., 2017, Sol. Phys., 292, 25

 R\"{u}diger, G., Hollerbach, R., 2004, The Magnetic
Universe: Geophysical and Astrophysical Dynamo Theory , pp. 343. ISBN
978-3-527-40409-4 (Wiley, July 2004)

 Sabbadin, F.,   Turatto, M.,   Cappellaro, E.,  S. Benetti, S.,   R. Ragazzoni R., 2004, A\&A, 

 Sabin, L., Zijlstra, A.A., Greaves, J.S., 2007, MNRAS, 376,
378

 Sabin, L., Wade, G.A.,  L\`{e}bre, A.,   2015, MNRAS 446, 1988

 Sahai, R. 2002, RevMexAA (SC), 13, 133

 Schwarz, H.,Aspin, C, Corradi, R.L.M., Reipurth, Bo., 1997,
A\&A, 319, 267

 Vainshtein, S.I.,  Rosner, R., 1991, ApJ, 376, 199

  Vlemmings, W.H.T., van Langevelde, H.J.,  Diamond, P.J., 2005, A\&A,  434, 1029

 Vlemmings, W. H. T. 2012, in IAU Symp. 283, Planetary Nebulae: An Eye to
the Future, ed. A. Machado, L. Stanghellini and  D. Sch\"{o}nberner (Cambridge:
Cambridge Univ. Press), 176

\end{document}